\documentclass{svjour3}   
\usepackage{graphicx}
\usepackage{amssymb,amsmath}
\usepackage[colon,authoryear]{natbib}
\begin{document}

\title{The rectilinear three body problem as a basis for studying highly-eccentric systems}

\author{G. Voyatzis, K. Tsiganis, M. Gaitanas}

\institute{Department of Physics, Aristotle University of Thessaloniki,\at 54124,
              Thessaloniki,\at Greece \\
              \email{voyatzis@auth.gr, tsiganis@auth.gr, mgaitana@physics.auth.gr} }    
\maketitle

\begin{abstract}
The rectilinear elliptic restricted Three Body Problem (TBP) is the limiting case of the elliptic restricted TBP when the motion of the primaries is described by a Keplerian ellipse with eccentricity $e'=1$, but the collision of the primaries is assumed to be a non-singular point. The rectilinear model has been proposed as a starting model for studying the dynamics of motion around highly eccentric binary systems. Broucke (1969) explored the rectilinear problem and obtained isolated periodic orbits for mass parameter $\mu=0.5$ (equal masses of the primaries). We found that all orbits obtained by Broucke are linearly unstable. We extend Broucke's computations by using a finer search for symmetric periodic orbits and computing their linear stability. We found a large number of periodic orbits, but only eight of them were found to be linearly stable and are associated with particular mean motion resonances. These stable orbits are used as generating orbits for continuation with respect to $\mu$ and $e'<1$. Also, continuation of periodic solutions with respect to the mass of the small body can be applied by using the general TBP. FLI maps of dynamical stability show that stable periodic orbits are surrounded in phase space with regions of regular orbits indicating that systems of very highly eccentric orbits can be found in stable resonant configurations. As an application we present a stability study for the planetary system HD7449.   
\end{abstract}

\section{Introduction}
 
The  planar rectilinear elliptic restricted three body problem or, simply, {\em rectilinear problem}, is a special case of the classical elliptic restricted three body problem (ERTBP) where the primaries oscillate on a straight line following Kepler's equation. Namely, we assume that the the primaries move on an ellipse of semimajor axis $a'$ and eccentricity equal to unity ($e'=1$). It was studied by \cite{schubart56} who obtained a rectilinear periodic orbit of the small body which is vertical to the line of equal mass primaries and in synchronization with them. This periodic orbit, which can be considered as a special orbit of the Sitnikov's problem, too, has  been suggested as a starting orbit for computing periodic orbits in the ERTBP. Indeed,  Schubart's periodic solution has been continued by \cite{broucke69} with respect to the mass ratio and the eccentricity of the primaries. In the paper of Broucke, which in the following will be referred to as {\em paper-I}, some new periodic orbits have been computed for the case of equal mass primaries. Our computations showed that all the above mentioned periodic solutions are unstable and cannot describe potential satellite or planetary orbits. In this paper we focus our study on determining stable periodic orbits and the phase space domain of regular orbits.

Although the rectilinear model is a toy model of the TBP, like Sitnikov's or the two-fixed centers model, it can contribute in understanding the dynamics of the motion of small bodies around highly eccentric primaries. This may include circumstellar or circumbinary motion of dust in disks around eccentric binary stars \citep{pichardo08} or planetary motion in very eccentric binary systems and stability criteria \citep{dvorak02,barnes06}. Also in the low mass ratio limit of the primaries we can study planetary systems consisting of a highly eccentric massive planet \citep{av16}.     

Following paper-I, we define the ERTBP by considering two primaries, $P_1$ and $P_2$, of mass $m_1=1-\mu$ and $m_2=\mu$ ($\mu<0.5$), respectively, and with relative elliptic motion of eccentricity $e'$, semimajor axis $a'=1$, period $T'=2\pi$ and the line of apsides coincides with the inertial $Ox$ axis ($\varpi'=0$). Then, by using the eccentric anomaly $E$, their distance is given by
\begin{equation} \label{rdistgen}
r'=1-e'\cos E,
\end{equation}
and their position $\mathbf{r}_i=(x_i,y_i)$ in the inertial $Oxy$ frame by the equations
\begin{equation} \label{xyposgen}  
\begin{array}{lll} 
x_1=-\mu(\cos E -e'), & \quad & y_1=-\mu \sqrt{1-e'^2} \sin E \\
x_2=(1-\mu)(\cos E -e'), & \quad & y_2=(1-\mu)\sqrt{1-e'^2} \sin E.
\end{array}
\end{equation}  
The position $\mathbf{r}=(x,y)$ of a third massless body, which interacts gravitationally with the primaries, is given by the equation of motion
\begin{equation} \label{ODEs}
 \ddot{\mathbf{r}} = -(1-\mu)\dfrac{\mathbf{r}-\mathbf{r_1}}{r_{01}^3}-\mu \dfrac{\mathbf{r}-\mathbf{r_2}}{r_{02}^3},
\end{equation}
where $r_{0i}^2=(x-x_i)^2+(y-y_i)^2$. Eq. (\ref{ODEs}) obeys the three body problem symmetry $\Sigma : (t,x,y) \rightarrow (-t,x,-y)$. Also, for $\mu=1/2$ the symmetry $\Sigma' : (t,-x,y) \rightarrow (t,x,y)$ is valid. 

The manipulation of the system (\ref{ODEs}) requires the relation between the eccentric anomaly and time, which is not given in closed form but it is determined through Kepler's equation
\begin{equation}\label{KeplerEq}
t-\tau=E-e'\sin E,
\end{equation}
where $\tau$ indicates the pericenter passage and $M'=t-\tau$ is the mean anomaly. 

\begin{figure}
\centering
\includegraphics[width=0.95\textwidth]{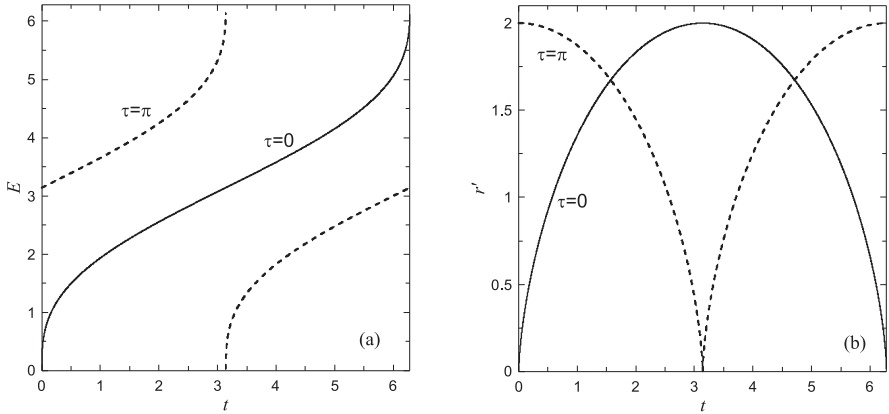}
\caption{{\bf a.} The solution $E=E(t)$ of Kepler's equation for $e'=1$  {\bf b.}  The relative distance of primaries $r'=r'(t)$ for $e'=1$.} 
\label{FigKeplerSol}
\end{figure}  

The rectilinear problem is derived directly from the above equations by setting $e'=1$. Then the primaries move on the $Ox$ axis ($y_1=y_2=0$) and
\begin{equation} \label{xposrect}
x_1=\mu(1-\cos E),\quad x_2=-(1-\mu)(1-\cos E).
\end{equation}
Thus, when $M'=0$ the primaries are located both at $O$ (periapsis) while for $M'=\pi$ we get $x_1=2\mu>0$ and $x_2=-2(1-\mu)<0$ (apoapsis). Equation (\ref{KeplerEq}) can be solved efficiently with a Newton-Raphson method and by using as initial guess value \citep{danby87} 
$$
E_0=t-\tau+0.85 \textnormal{sign} \left(\sin(t-\tau) \right ) e'.
$$            
The above value is sufficient for $e'=1$, too. Our code provides accuracy better than $10^{-15}$ checking always the Newton-Raphson convergence.  For the rectilinear case the functions $E(t)$ and $r'(t)$ are presented in fig. \ref{FigKeplerSol}. The ODEs (\ref{ODEs}) are numerically integrated by using the Bulirsch -- Stoer algorithm with prescribed accuracy $\geq 10^{-12}$.

\section{Bounded and escape orbits}
\begin{figure}
\centering
\includegraphics[width=0.95\textwidth]{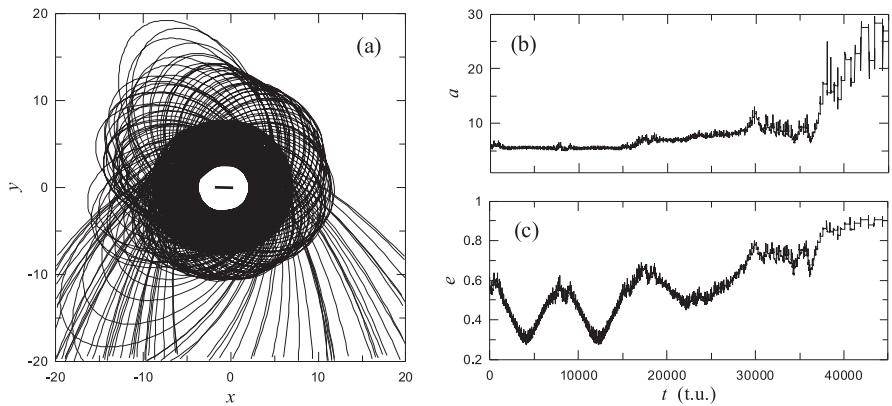}
\caption{An example of an escape orbit that avoids close encounter ($x_0=2.485$, $\dot{y}_0=0.8$, $\tau=0$, $\mu=0.1$). {\bf a.} the orbit in the $xy$ plane. The line segment indicates the interval of motion of the primaries. {\bf b.} the evolution of the semimajor axis computed in the barycentric frame. {\bf c.} the evolution of the eccentricity.}
\label{FigDiffOrb}
\end{figure}

In this section we present a preliminary study for determining initial conditions for bounded motion. Since the rectilinear model does not possess a Jacobi-like integral, we cannot obtain a Hill stability criterion. When the small body is quite far from the origin, we can use the two body approximation and the escape criterion 
\begin{equation} \label{Vesc}
\dot x^2+\dot y^2 \geq \frac{2}{r},
\end{equation}
In the three body problem, the typical condition for the ejection of the small body is a sequence of close encounters. This is also a typical case for the rectilinear model. But apart from close encounters, the numerical simulations showed escape through chaotic diffusion. In Fig. \ref{FigDiffOrb} we present an example of such a diffusive orbit in the absence of close encounters. We clearly observe an almost circular domain around the origin, which encompasses the interval of motion of the primaries. After some time, the condition (\ref{Vesc}) holds and an almost parabolic orbit is obtained.    

The simplicity of the equations of motion permit us to perform a large amount of numerical integrations for determining the initial conditions for bounded motion. We consider grids of initial conditions on the plane $(x_0,\dot y_0)$ and $y_0=\dot x_0=0$ and we numerically integrate the orbits for $N_{max}=10^5$ periods of the primaries. We consider that an orbit, which starts with $r\leq 5$, is unbounded when we obtain that $r>30$ at an iteration $N_{esc}$. By assigning a color map to the value $N_{esc}$, we construct {\em escape time maps}. 

\begin{figure}
\centering
\includegraphics[width=0.95\textwidth]{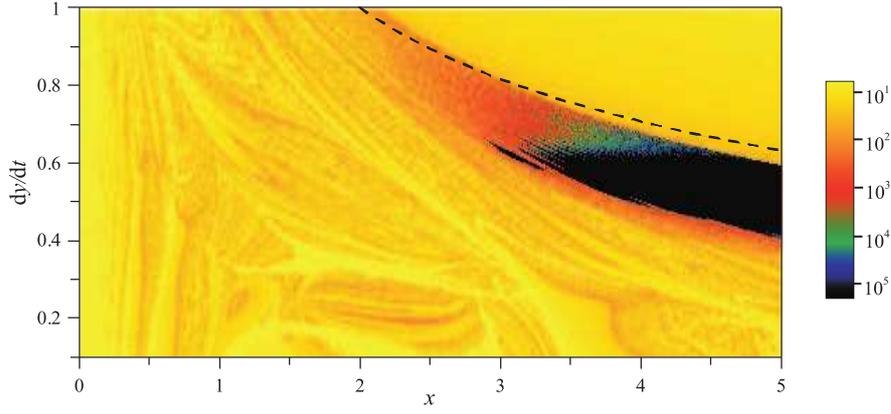}
\caption{Escape time map of size $500\times 200$ for $\mu=0.5$ and $\tau=0$. The color scale indicates the number of periods of the primaries ($N_{esc}$) where the escape condition becomes true. Black color indicates bounded orbits (at least for $10^5$ iterations). The dashed line indicates the approximate escape velocity given by inequality (\ref{Vesc}). }
\label{FigEscmap1}
\end{figure}

In Fig. \ref{FigEscmap1} we present the escape time map for $\mu=0.5$ and $\tau=0$. We remark that due to the symmetries $\Sigma$ and $\Sigma'$ the map is the same for the other quadrants of the $Oxy$ plane. The map shows the existence of bounded motion (dark area) when the small body starts quite far from the primaries, particularly for $x>2.92$ and $\dot y <0.66$. All bounded orbits are circumbinary orbits with (approximate) period of revolution $T_r>10\pi$. Approximately, this area is delimited from above by the escape condition (\ref{Vesc}) and seems to extend for $x\rightarrow \infty$ where the system becomes hierarchical and can be approximated by the two body problem. Generally, by computing maps of higher resolution or by zooming in particular areas in phase space, we observe fractal structures between stable and escape domains. This is a common feature in open systems of celestial mechanics and has been observed in the Sitnikov problem \citep{Kovacks09}, the Trojan problem \citep{Paez2015}, the Earth-Moon system \citep{Assis14} and in galactic models \citep{Contopoulos12}.        

\begin{figure}
$
\begin{array}{ccc}
\includegraphics[width=0.45\textwidth]{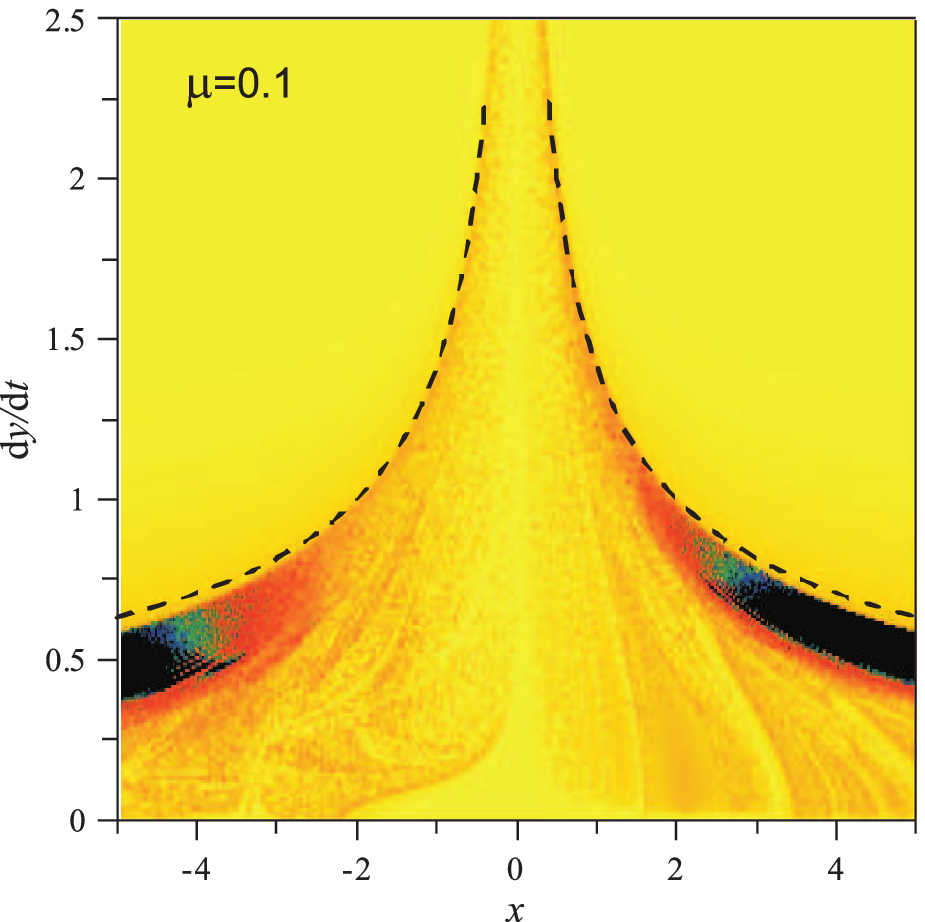} & \quad & \includegraphics[width=0.45\textwidth]{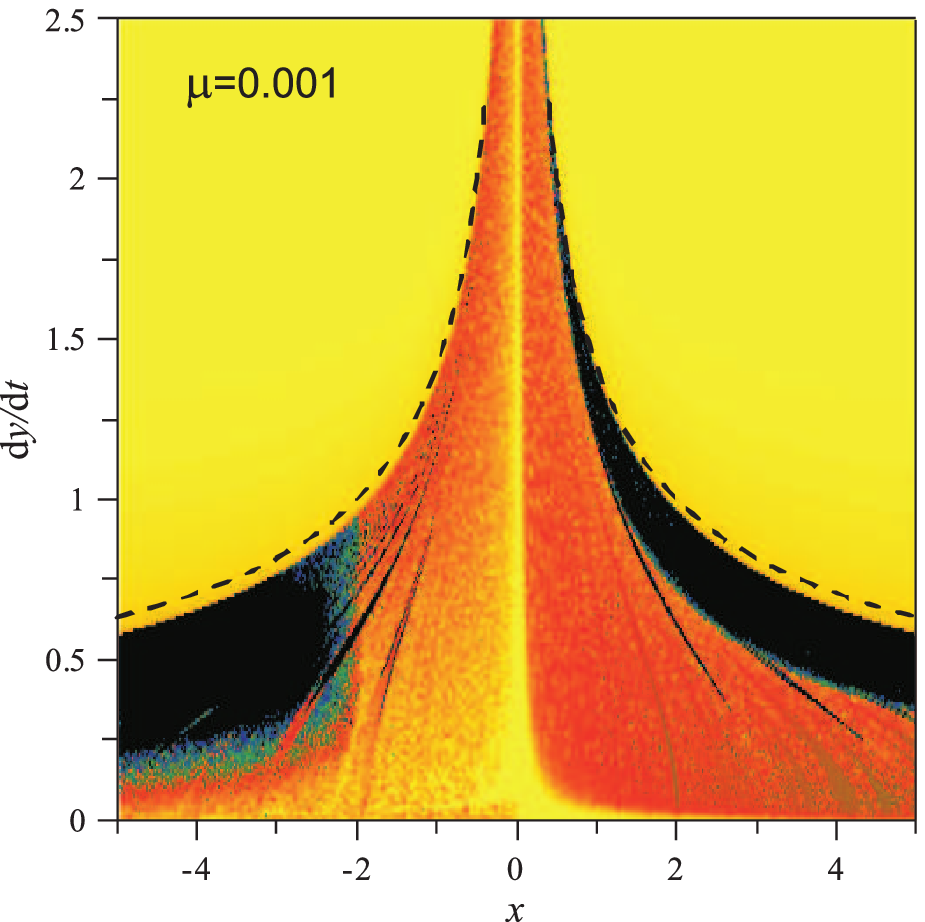} \\
\textnormal{(a)} & \quad & \textnormal{(b)}  
\end{array}$
\caption{Escape time map of size $500\times 250$ for {\bf a.} $\mu=0.1$, $\tau=0$  {\bf b.} $\mu=0.001$, $\tau=0$. The color scale is the same as in Fig. \ref{FigEscmap1}. The dashed line indicates the approximate escape velocity.}
\label{FigEscmap2}
\end{figure}

In Fig. \ref{FigEscmap2}a we present the escape time map for $\mu=0.1$ and $\tau=0$. Qualitatively we obtain the same picture with that for $\mu=0.5$. However, now the symmetry $\Sigma'$, with respect to the axis $x=0$,  does not hold. The larger primary $P_1$ moves in the domain $0 \leq x\leq 0.2$ and the lighter one, $P_2$, in $-1.8\leq x\leq 0$. Bounded motion is obtained in the dark regions of the map appeared in the intervals $x>2.38$ and $x<-3.55$. 

\begin{figure}
$
\begin{array}{ccc}
\includegraphics[width=0.48\textwidth]{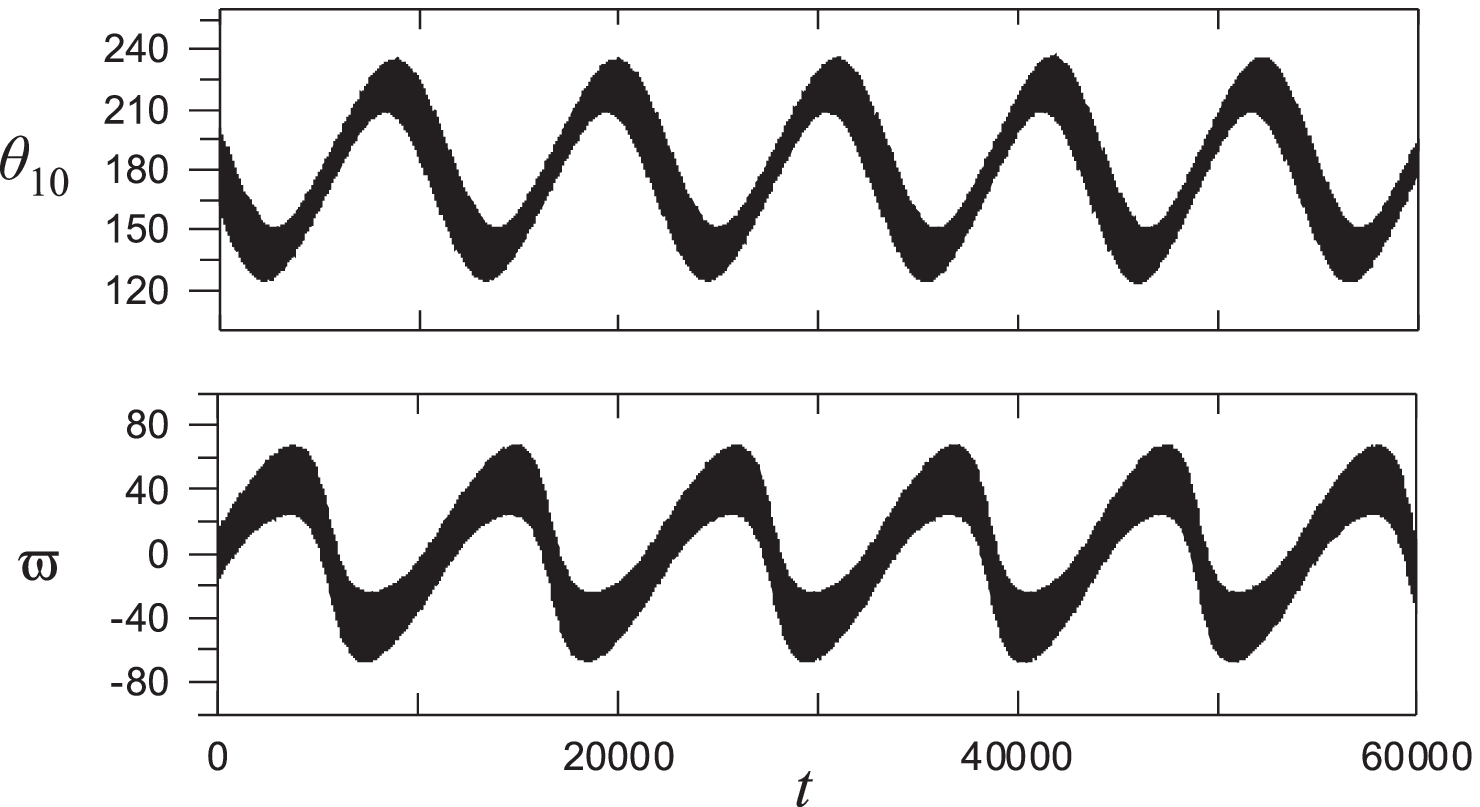} &\quad & \includegraphics[width=0.48\textwidth]{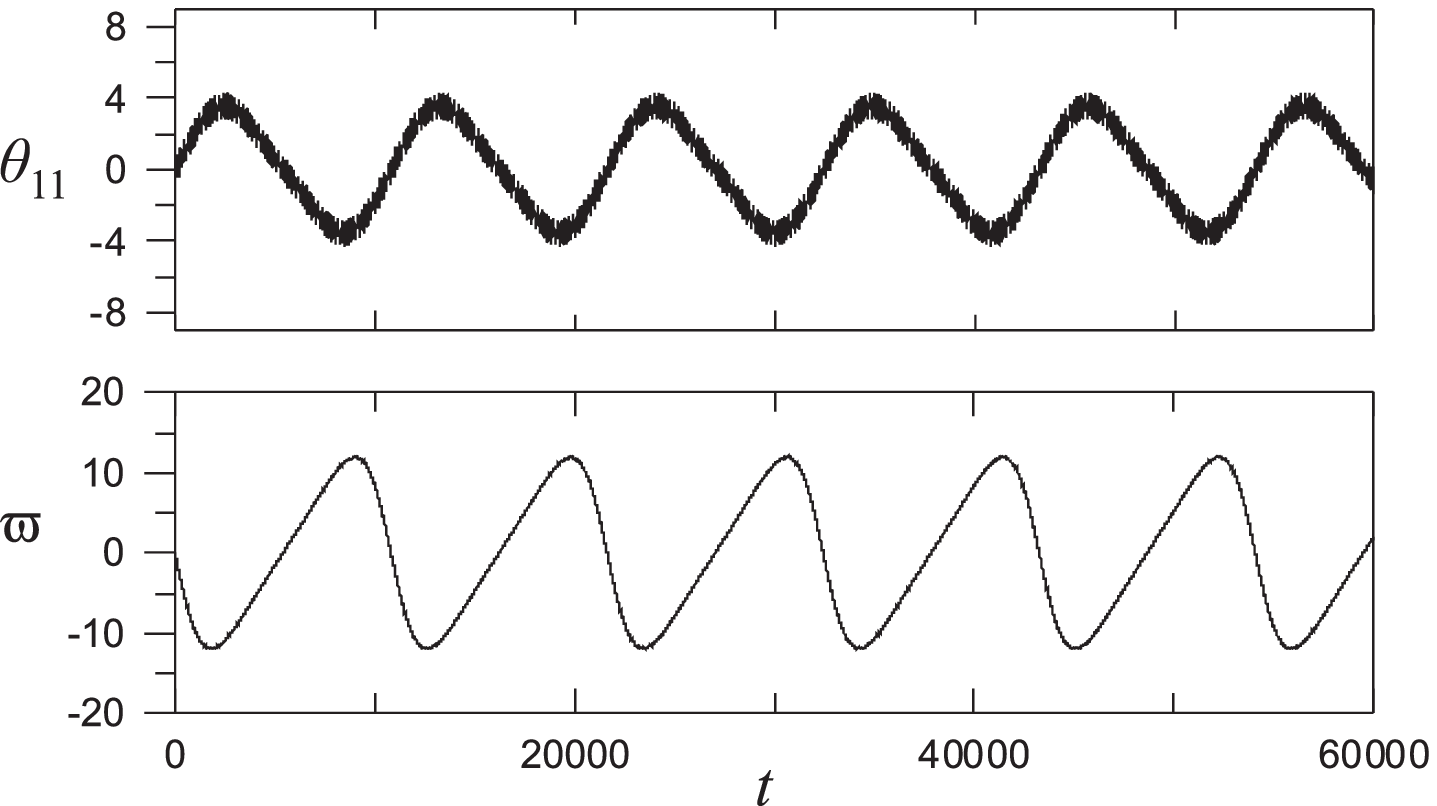} \\
\textnormal{(a)} &\quad & \textnormal{(b)}  
\end{array}
$
\caption{Librations of the resonant angle $\theta_{pq}$ and the longitude of pericenter $\varpi$ {\bf a.} $p=1, q=0$, for initial conditions $x_0=-1.467$, $y_0=\dot x_0=0$, $\dot y_0=0.587$, $\mu=0.001$  {\bf b.} $p=1, q=1$, $x_0=1.467$, $y_0=\dot x_0=0$, $\dot y_0=0.856$, $\mu=0.001$.}
\label{FigLib}
\end{figure} 

The escape time map for $\mu=0.001$ is presented in Fig. \ref{FigEscmap2}b. We can observe a wider region of bounded motion with respect to the case of larger $\mu$ values. For $x>0$, where the larger primary moves, we have initial conditions of bounded motion for relatively large values of $\dot y$ and close to the escape boundary. Again, the majority of bounded orbits are circumbinary but there exist orbits of satellite type too, as we will show in the next section. E.g. in the thin tangles at $(x,\dot y)\approx (-1.5,0.6)$, the orbits revolve around the larger primary $P_1$ (which is now located almost at the origin) with an approximate revolution period $T_r\approx 2 \pi$ avoiding the collision of the smaller primary. Such satellite orbits are in $1:1$ resonance and the synodic angle $\theta_{10}=\lambda-\lambda'$, where $\lambda$ is the mean longitude of the small body around the origin and $\lambda'=M'$ (since $\varpi'=0$), librates as it is shown in Fig. \ref{FigLib}a \footnote{We note that the orbital elements for the rectilinear model are computed relatively to the barycenter of the primaries}. Generally, the orbits with initial conditions in the tangles shown in the escape time map are resonant, $T_r\approx 2\pi\frac{p+q}{p}$, $p,q\in \mathbb{N}$, and the angle 
\begin{equation}\label{EqResAngle}
\theta_{pq}=(p+q)\lambda-p \lambda'
\end{equation}
librates. An example of a libration along a 2:1 circumbinary resonant orbit is shown in Fig. \ref{FigLib}b. The longitude of pericenter librates too, either around $0$ or $\pi$. We mention that librations of resonant angles have been found in \cite{schubart17} for planets around elliptic binary systems.

\section{Symmetric periodic orbits} \label{secSPO}
Resonant domains of motion in phase space are associated with periodic solutions. Since the system is periodic of period $2\pi$, periodic orbits must have period $T=2k\pi$, where $k$ is an integer called {\em period multiplicity}. Due to the symmetry $\Sigma$, symmetric periodic orbits can be obtained for initial conditions 
\begin{equation}
x(0)=x_0,\; \dot y(0)=\dot y_0\quad (y(0)=\dot x(0)=0)
\end{equation}
when the primaries are at periapsis ($\tau=0$) or apoapsis ($\tau=\pi$). Such initial conditions correspond to a symmetric periodic orbit if the periodicity condition 
\begin{equation} \label{EqPCON}
 y(\pi)=\dot x(\pi)=0
\end{equation}
is satisfied. This means that a periodic orbit shows two perpendicular crossings with the $y=0$ axis, at $t=0$ or $T$ and at $t=T/2$. At these moments the primaries are at periapsis or apoapsis. The points ($x_0,\dot y_0$), which correspond to periodic orbits, are isolated in the plane $x-\dot y$. Also, if ($x_0,\dot y_0$) corresponds to a periodic orbit, then ($x_0,-\dot y_0$) corresponds to the same periodic orbit. Details about the computation of periodic solutions and their linear stability can be found in paper-I and in \citet{gv17}.         

In the present work, we performed a thorough computational search for periodic orbits in the rectilinear problem. Using grids of initial conditions, similar to those of the escape time maps presented in the previous section, we perform differential corrections in order to succeed in satisfying the periodicity conditions (\ref{EqPCON}) with a high precision. Their linear stability is also computed according to the method described in paper-I. There, some orbits of the rectilinear model  have been computed but the study of their stability is missing. Our study showed that these periodic orbits are unstable. So our challenge is to find linearly stable periodic orbits, which should verify the existence of stable motion when the primaries revolve in highly eccentric orbits. We restrict our study for relatively small periods, particularly for period multiplicity $1\leq k \leq 4$. For $k$=odd, the choice of the initial position of the primaries, namely $\tau=0$ or $\tau=\pi$, is not essential since both cases can represent the same periodic orbit. However, for $k$=even, both initial positions of the primaries should be considered in order to compute potentially all periodic orbits. We studied the cases $\mu=0.5$, $0.1$, $0.01$ and $0.001$, but we present here results only for the cases $\mu=0.5$ and $0.001$. 

\begin{figure}
\centering
\includegraphics[width=0.95\textwidth]{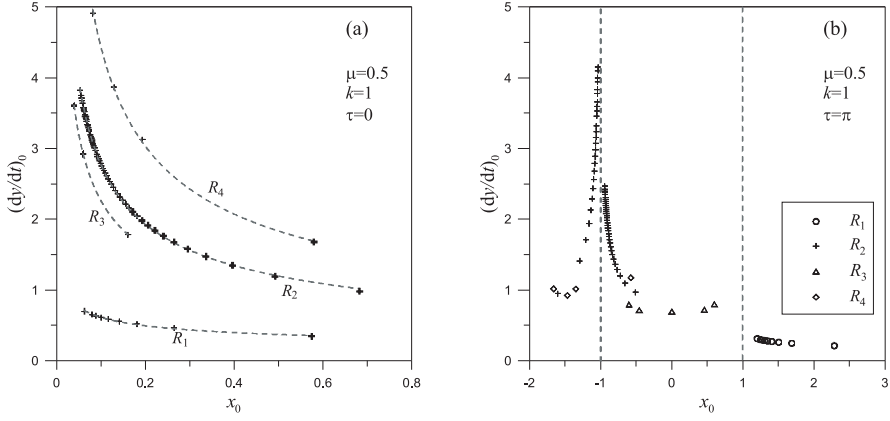}
\caption{Initial conditions of symmetric periodic orbits for $\mu=0.5$ and $k=1$. {\bf a.} initial conditions when the primaries are at periapsis (first perpendicular crossing of the orbits with the axis $y=0$). The dashed lines correspond to power law curves $\dot y_0\sim x_0^{-1/2}$. The bold crosses represent the orbits found also in paper-I  {\bf b.} initial conditions of the same orbits as in the left panel when the primaries are at apoapsis (second perpendicular crossing of the orbits with the axis $y=0$). The vertical dashed lines indicate the initial position of the primaries.}
\label{FigPomu05k1}
\end{figure}  

\begin{figure}
\centering
\includegraphics[width=0.95\textwidth]{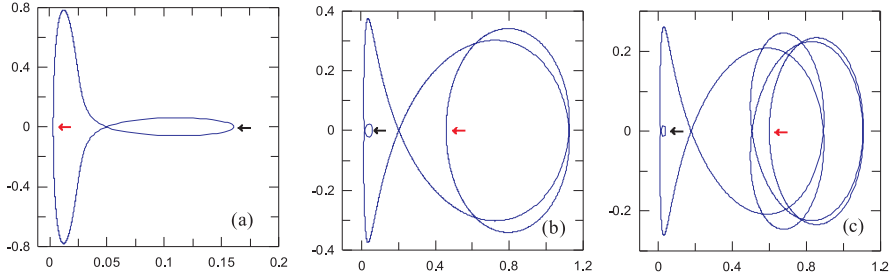}
\caption{Three samples of periodic orbits of the set $R_3$ presented on the inertial plane $Oxy$. The black arrow indicates the initial position when the primaries are at periapsis. The red arrow indicates the second perpendicular crossing at $t=T/2=\pi$, when the primaries are at apoapsis. }
\label{FigPosR3}
\end{figure}  

\subsection{The case $\mu=0.5$}
For equal primaries, our search can be restricted in the domain $x>0$, $\dot y>0$. If ($x_0, \dot y_0$) corresponds to a periodic orbit, then ($-x_0,\dot y_0$) corresponds to an other periodic orbit, which is the mirror image of the first one. In Fig. \ref{FigPomu05k1} we present the initial conditions of the periodic orbits with period $T=2\pi$ ($k=1$). In the left (right) panel the initial conditions correspond for $\tau=0$ ($\tau=\pi$). In the left panel the initial conditions correspond to the initial perpendicular crossing for $\tau=0$, while the right panel corresponds to the conditions at the second perpendicular crossing of the orbit with the axis $y=0$, where we can assume $t_0=0$ and $\tau=\pi$. From their distribution in the left panel we can classify the orbits in four different sets, $R_i$, $i=1,..,4$.  For each set the orbits are located on power law curves, $\dot y_0=c x_0^{-b}$,  with $b\approx 0.5$. The initial conditions of the  orbits of the set $R_2$ for $\tau=\pi$ are distributed in two curves that also follow a power law, one is located on the left of the primary $P_2$ with $b\approx 0.53$ and the other on the right with $b\approx 0.46$. The first orbit of each set (namely the orbit with the maximum $x_0$ in panel (a)) has a relatively simple geometrical shape but the orbits with $x_0\rightarrow 0$ are quite complicated showing many revolutions around one of the primaries. In paper-I, only the first orbits of the sets $R_1$ and $R_4$ have been found and many orbits of the set $R_2$.  Three of the five orbits of the set $R_3$ are presented on the $Oxy$ plane in Fig. \ref{FigPosR3}.  All these periodic orbits are linearly unstable and are located in the escape regime of the map shown in Fig. \ref{FigEscmap1}.         

\begin{figure}
\centering
\includegraphics[width=0.95\textwidth]{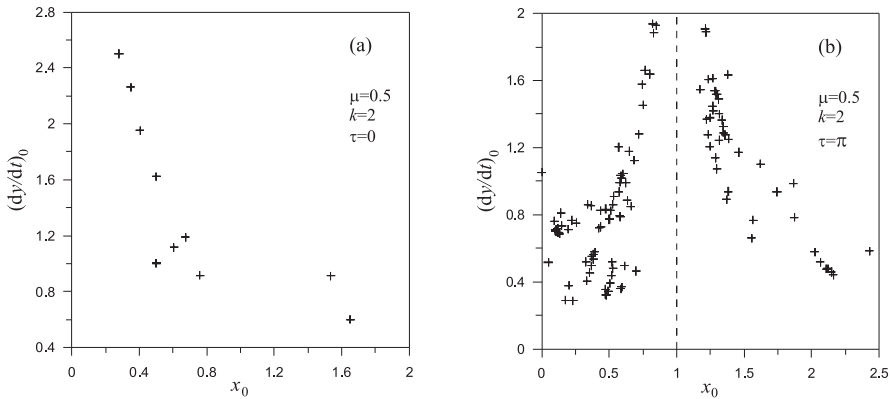}
\caption{Initial conditions ($x_0>0$) of symmetric periodic orbits for $\mu=0.5$ and $k=2$ and  {\bf a.} $\tau=0$ {\bf b.} $\tau=\pi$. }
\label{FigPomu05k2}
\end{figure}  

\begin{figure}
\centering
\includegraphics[width=0.95\textwidth]{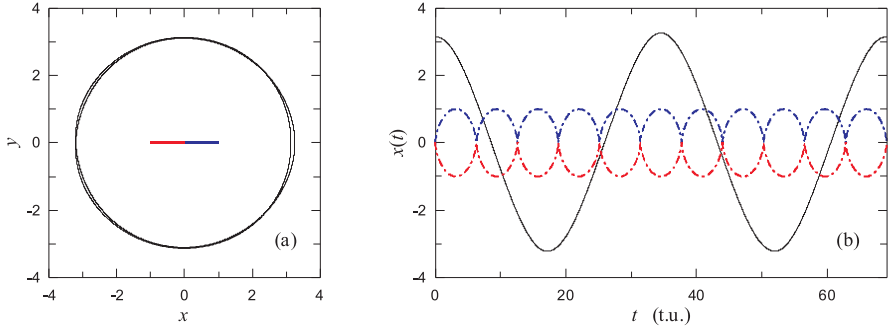}
\caption{The $11:2$ resonant stable periodic orbit ($T=22\pi$) {\bf a.} the orbit on the $Oxy$ plane and the interval of motion of the primaries {\bf b.} The $x=x(t)$ evolution of the primaries (dashed lines) and the small body (solid line) for one period. }
\label{FigPok11}
\end{figure}

In Fig. \ref{FigPomu05k2} we present the initial conditions for periodic orbits of period $T=4\pi$. Now the two panels present different periodic orbits. For $\tau=0$ the number of periodic orbits is quite smaller in comparison with the case $k=1$. For $\tau=\pi$ a significantly larger number of periodic orbits is detected. All these periodic orbits are unstable and the same case holds also for the orbits we found for $k=3$ and $k=4$. Nevertheless we cannot exclude the existence of stable periodic orbits for larger periods. E.g. by examining the escape time map for $\mu=0.5$ (see Fig. \ref{FigEscmap1}) we can observe an isolated island of stability at $(x,\dot y)\approx (3.2,0.6)$. Searching in this region for a periodic orbit of larger period multiplicity we found for $\tau=0$ a stable periodic orbit at $x_0$=$3.1410325550$, $\dot y_0$=$0.6112831375$ for period multiplicity $k=11$. This is a $11:2$ resonant circumbinary orbit and is presented in Fig. \ref{FigPok11}. 

\subsection{The case $\mu=0.001$}
This case may be thought of as a system where the primaries are a star (of the mass of the Sun) and a planet (of Jupiter's mass). The small body can be either a planet around the Sun (circumstellar type orbit) or it can revolve in an orbit outside of both planets (circumbinary type orbit). Other type of orbits, e.g. satellite orbit around the Jupiter, are found to be very unstable in the rectilinear model.   

\begin{figure}
\centering
\includegraphics[width=0.95\textwidth]{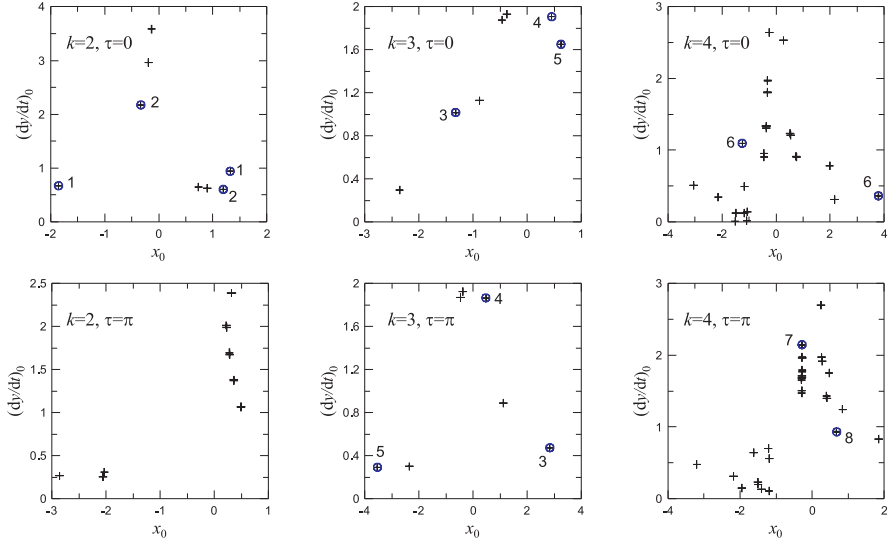}
\caption{Initial conditions of symmetric periodic orbits for $\mu=0.001$. The period multiplicity, $k$, and the pericenter passage, $\tau$, are indicated. The encircled crosses indicate the stable periodic orbits, which are numbered and are displayed twice, namely for initial conditions at $t=0$ and at $t=T/2$. The initial conditions at $t=0$  and $t=T/2$ of the orbits 7 and 8 almost coincide, so they seem to be displayed in a single position.}
\label{FigPomu001k}
\end{figure}  

\begin{table}
\caption{Initial conditions, period, orbital data and type of the 8 stable periodic orbits found for $\mu=0.001$. For all cases $\tau=0$ except for the orbits 7 and 8 where $\tau=\pi$.} \resizebox{\columnwidth}{!}
{
\begin{tabular}{cccccccccc}
\hline
No & $x_0$ & $\dot y_0$ & $T$ & $n/n'$ & $a$ & $e$ & $\varpi$ & $M$ & type\\ 
\hline
1  & 1.3210127289 & 0.9405671047  & $4\pi$ & 1:2 & 1.589 & 0.166 & 0 & $\pi$ & CS \\ 
2  & 1.1944758137 & 0.6035681942  & $4\pi$ & 3:2 & 0.760 & 0.565 & $\pi$ & $\pi$ & CS \\
3  & -1.3217481552& 1.0164547131  & $6\pi$ & 1:3 & 2.080 & 0.366 & $\pi$ & 0 & CS \\ 
4  & 0.4544892632 & 1.9070809129 & $6\pi$ & 2:3 & 1.310 & 0.653 & 0 & 0 & CB \\
5  & 0.6242478246 & 1.6505634562 & $6\pi$ & 1:3 & 2.080 & 0.700 & 0 & 0 & CB \\
6  & 3.7857752447 & 0.3626541268 & $8\pi$ & 1:4 & 2.525 & 0.503 & $\pi$ & 0 & CS \\
7* & -0.2788831282 & 2.1473648829 & $8\pi$ & 4:1 & 0.386 & 0.270 & $\pi$ & 0 & CS \\
8* & 0.6819941811 &  0.9313863887 & $8\pi$ & 3:1 & 0.481 & 0.404 & $\pi$ & $\pi$ & CS \\ 
\hline
\end{tabular}
}
\label{tab}
\end{table}

Our search for symmetric periodic orbits determined 10 periodic orbits of single period multiplicity ($k=1$), which are linearly unstable. For $k=2$, $3$ and $4$ we found the periodic orbits with initial conditions presented in Fig. \ref{FigPomu001k}. Each periodic orbit is depicted by two crosses, one corresponds to initial conditions for $t=0$ and the other to $t=T/2$. In the case of odd period multiplicity ($k=3$) the panels for $\tau=0$ and $\tau=\pi$ present the same orbits but at initial conditions of different phase. The encircled crosses, which are numbered, indicate linearly stable orbits. All these orbits are almost elliptical and can be described by the osculating orbital elements $a$, $e$, $\varpi$ and $M$. Also the periodic orbits are mean motion resonant and such a resonance is given by the ratio $n/n'$, where $n'=1$ is the mean motion of the primaries and $n$ the mean motion of the massless body, which must be rational and it is either $n\geq 1$ or $n<1$.
   
The initial conditions and the orbital elements of the eight stable periodic orbits found are given in Table \ref{tab}. The orbits 1,3 and 6 are circumstellar (CS), but their approximate period of one revolution around the Sun is larger than $2\pi$. The orbit 2 is circumstellar, too, with revolution period $T_r\approx 4\pi/3$ ($<2\pi$). The orbits 4 and 5 are circumbinary (CB), namely the orbit encircles all the $x$-interval of the motion of the primaries and, certainly, the period of one revolution is larger than $2\pi$. The circumstellar orbits 7 and 8, which are the only stable orbits found for $\tau=\pi$, may be characterized as satellite orbits since the revolution around the Sun takes place in time, $T_r$, which is a sub-multiple of $2\pi$ and their initial conditions for $t=0$ and $t=T/2$ are almost the same. Furthermore, we can observe that most of the periodic orbits are of moderate or high eccentricity. 

The fact that the periodic orbits of Table \ref{tab} are linearly stable, permits us to conclude the existence of domains of stable (regular) motion in phase space in the neigbourhood of such periodic orbits. In order to verify such a conclusion we compute stability maps based on the computation of the fast Lyapunov indicator (\cite{froe97}) defined in the particular case as $FLI={\mathbf\xi}/t$, where ${\mathbf\xi}={\mathbf\xi}(t)$ the deviation vector computed by solving the linearized ODE's of the equations of motion (\ref{ODEs}). We depict the value of FLI at $t_{max}=10000$ periods of the primaries or we stop the integration when $FLI>10^{30}$. When the orbit escapes we stop the integration and set the value $FLI=10^{40}$, which is indicated in the maps by the lighter (yellow) color. Generally, values of $FLI<10^{5}$ indicate regular motion, which is depicted in the maps with the dark colors. The maps can be given in the plane of initial conditions ($x_0$,$\dot y_0$), with fixed $y_0=\dot x_0=0$, or in the plane of the initial orbital elements, $a$ and $e$ by fixing the initial $\varpi$ and $M$ at values given in table \ref{tab}.  

Fig. \ref{FigPoMaps} presents the stability maps around the stable periodic orbits 1,4 and 7, which are indicated by the cross. The panels in the top present the maps in the ($x_0$,$\dot y_0$) plane and panels in bottom present the maps in the ($a$,$e$) plane.  It is clearly observed that for each case, the stable periodic orbit is located almost in the middle of a domain of regular orbits. Such stability domains look like strips in the ($x_0$,$\dot y_0$) planes.  Chaotic orbits outside this domain, soon or far escape. In case of orbit 1, the stability domain is quite wide and includes regular orbits from low up to high eccentricities. In case of the circumbinary orbit 4 (panel b) the stability domain is restricted only in very high eccentricities. The stability domain is relatively very small in the case of the satellite orbit 7.

\begin{figure}
$
\begin{array}{c}
\includegraphics[width=0.95\textwidth]{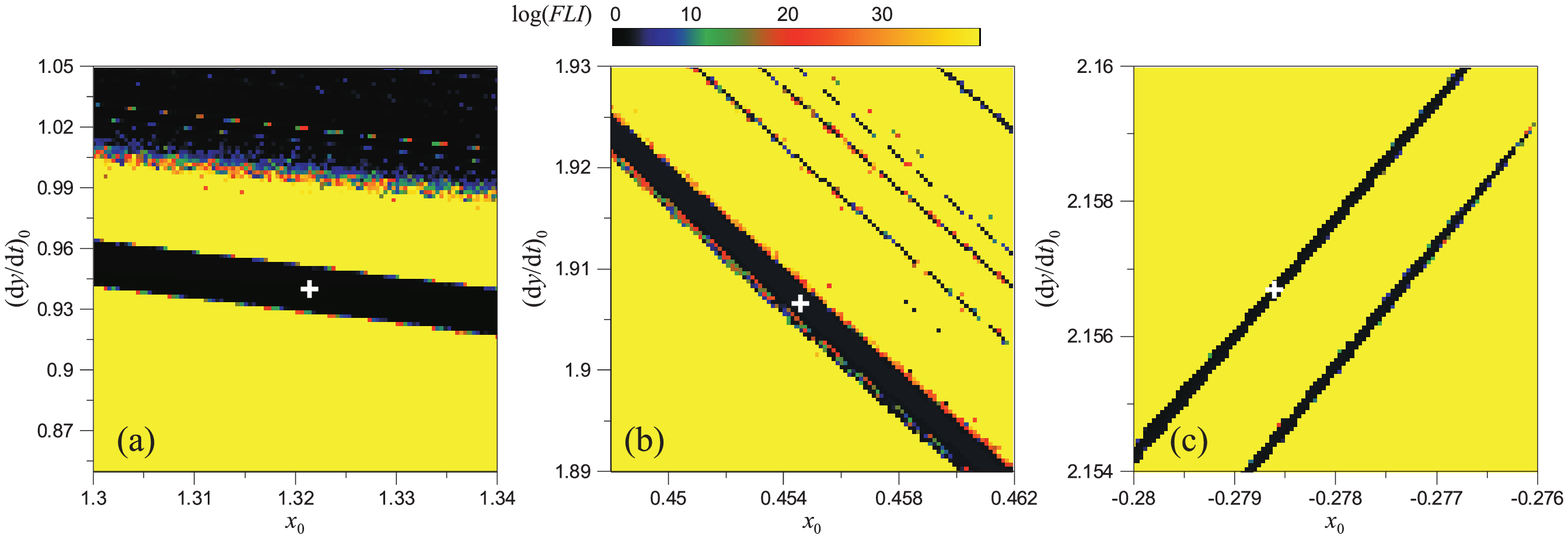} \\ \includegraphics[width=0.95\textwidth]{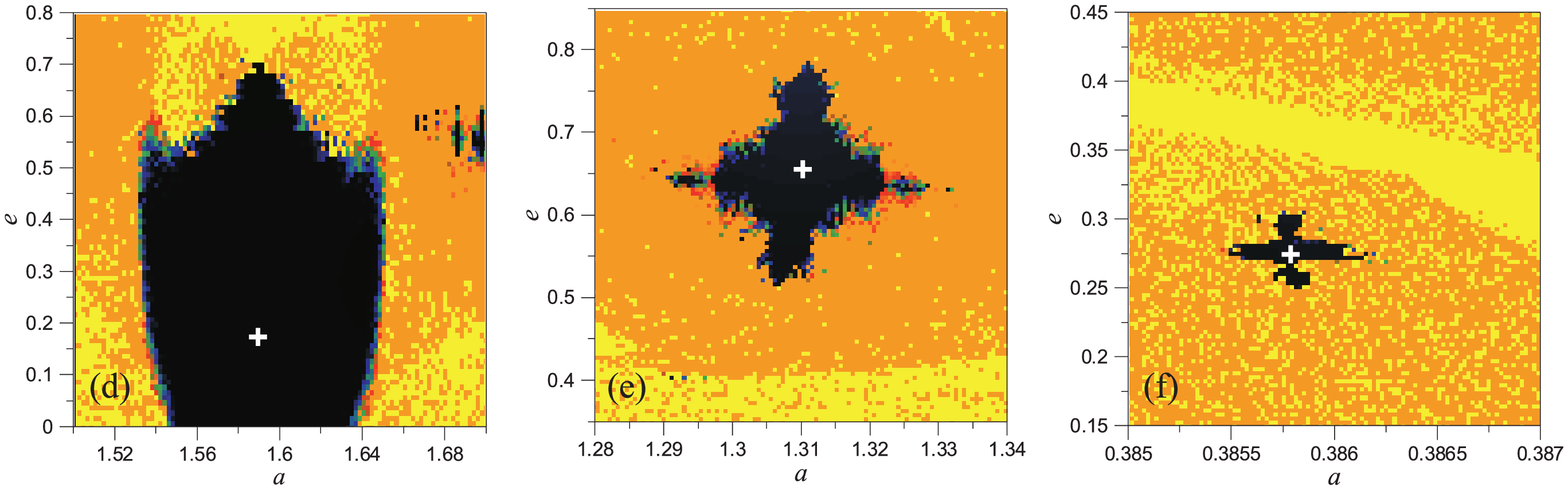}  
\end{array}$
\caption{Stability maps of size $100\times 100$ in the vicinity of periodic orbits 1, 4 and 7 given in the table \ref{tab} (panels (a,d), (b,e) and (c,f), respectively). Top row presents the maps in the ($x_0$,$\dot y_0$) plane. Bottom row presents the maps in the ($a$,$e$) plane. The initial values for the pericenter longitude $\varpi$ and the mean anomaly $M$ are those given in the table. The position of the stable periodic orbit is indicated by the cross. Dark regions indicate initial conditions for regular motion.}
\label{FigPoMaps}
\end{figure}

\section{Continuation of stable periodic orbits}
In the circular or the elliptic restricted TBP, continuation of periodic solutions with respect to the mass can be shown \citep[see e.g.,][and references therein]{ichkm78}. The initial and periodicity conditions discussed in section \ref{secSPO} are valid for the ERTBP with $0<e'\leq1$. Therefore, as it is shown also in paper-I,  periodic solutions of the rectilinear problem can be continued for $e'<1$. Subsequently, by varying the mass parameter $\mu$ or the eccentricity $e'$ we obtain continuous monoparametric sets of periodic solutions of constant period. We call them families $f_m$ and $f_e$ (or $f'_e$ if the orbits are retrograde), respectively. Along such families the stability indices of linear analysis vary and, therefore, along a family we may observe changes of the stability type. For continuity reasons, when a  stable (unstable) orbit is continued we obtain stable (unstable) orbits at least for small changes of the parameters $\mu$ or $e'$. In the following we present our results from the continuation method of the stable orbits shown in Table \ref{tab}. 

\begin{figure}
$
\begin{array}{ccc}
\includegraphics[width=0.48\textwidth]{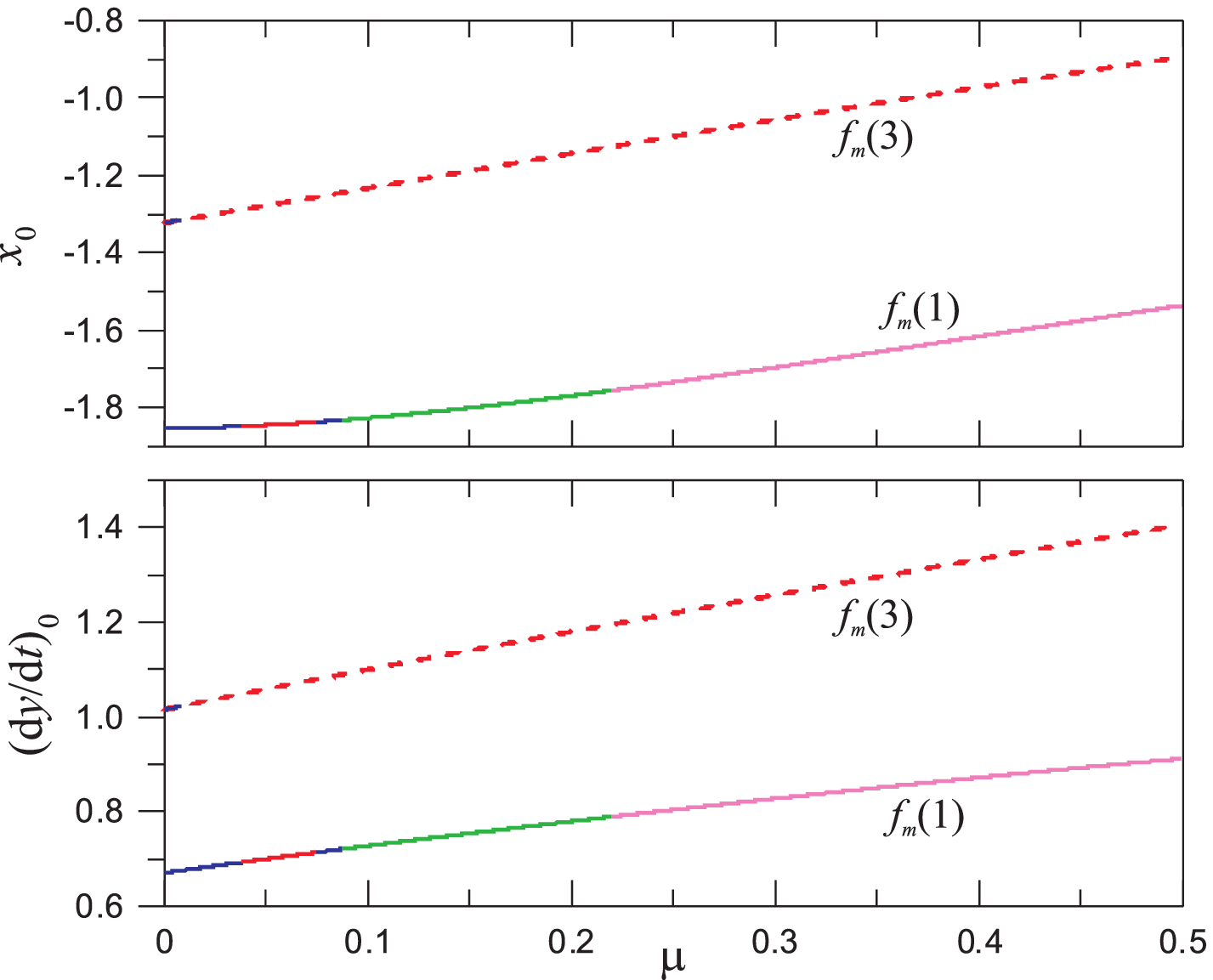} &\quad & \includegraphics[width=0.48\textwidth]{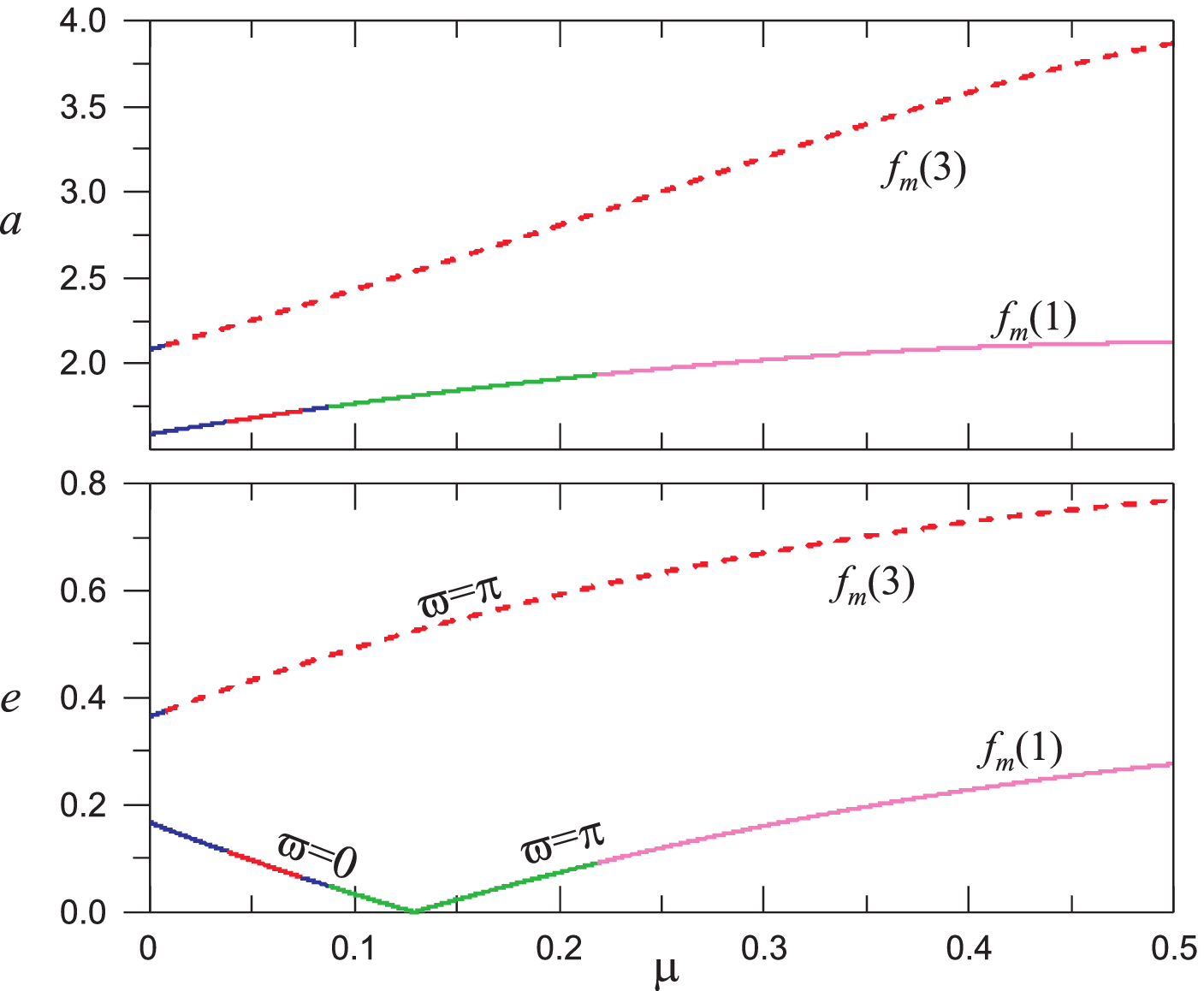} \\
\end{array}$
\caption{Continuation of periodic orbits 1 and 3 with respect to the mass parameter $\mu$ (families $f_m(1)$ and $f_m(3)$, respectively). The panels show the initial conditions and the orbital elements of the periodic orbits along the families. The blue line segments indicate that the orbits are linearly stable. Red, green and pink colors indicate single, complex and double instability, respectively. }
\label{FigCondm}
\end{figure} 

\subsection{Continuation with respect to the mass}
In figure Fig. \ref{FigCondm} we show the continuation of the periodic orbits 1 and 3 (see table \ref{tab}) by presenting the variation of initial conditions and orbital elements with respect to $\mu$. In both cases continuation is feasible up to $\mu=1$ but, since for $\mu>0.5$ the same orbits are obtained due to the symmetry of the system, we present the families up to $\mu=0.5$. For the family $f_m(1)$, generated from orbit 1, the particular initial conditions $x_0$ and $\dot{y}_0$ increase as $\mu$ increases and this is also the case for the semimajor axis computed with respect to the barycenter of the primaries at ($x_0$, $\dot{y}_0$). The eccentricity decreases and at $\mu=0.129$ the orbit becomes almost circular and the angle of apsides changes from $\varpi=0$ to $\varpi=\pi$. The family starts with a segment of stable periodic orbits in the interval $0\leq \mu<0.03811$. Then the orbits become single unstable and in the interval $0.0744 < \mu < 0.0873$ become stable again. After this interval we obtain complex unstable orbits which for $\mu>0.2192$ become doubly unstable. All these stability types are defined in paper-I and the mentioned stability changes are consistent with respect to the continuation. 

The family $f_m(3)$, generated from orbit 3, consists of stable orbits only for a small values of $\mu$, particularly for $\mu<0.0081$. After this value, the orbits continue as single unstable.

\begin{figure}
$
\begin{array}{ccc}
\includegraphics[width=0.48\textwidth]{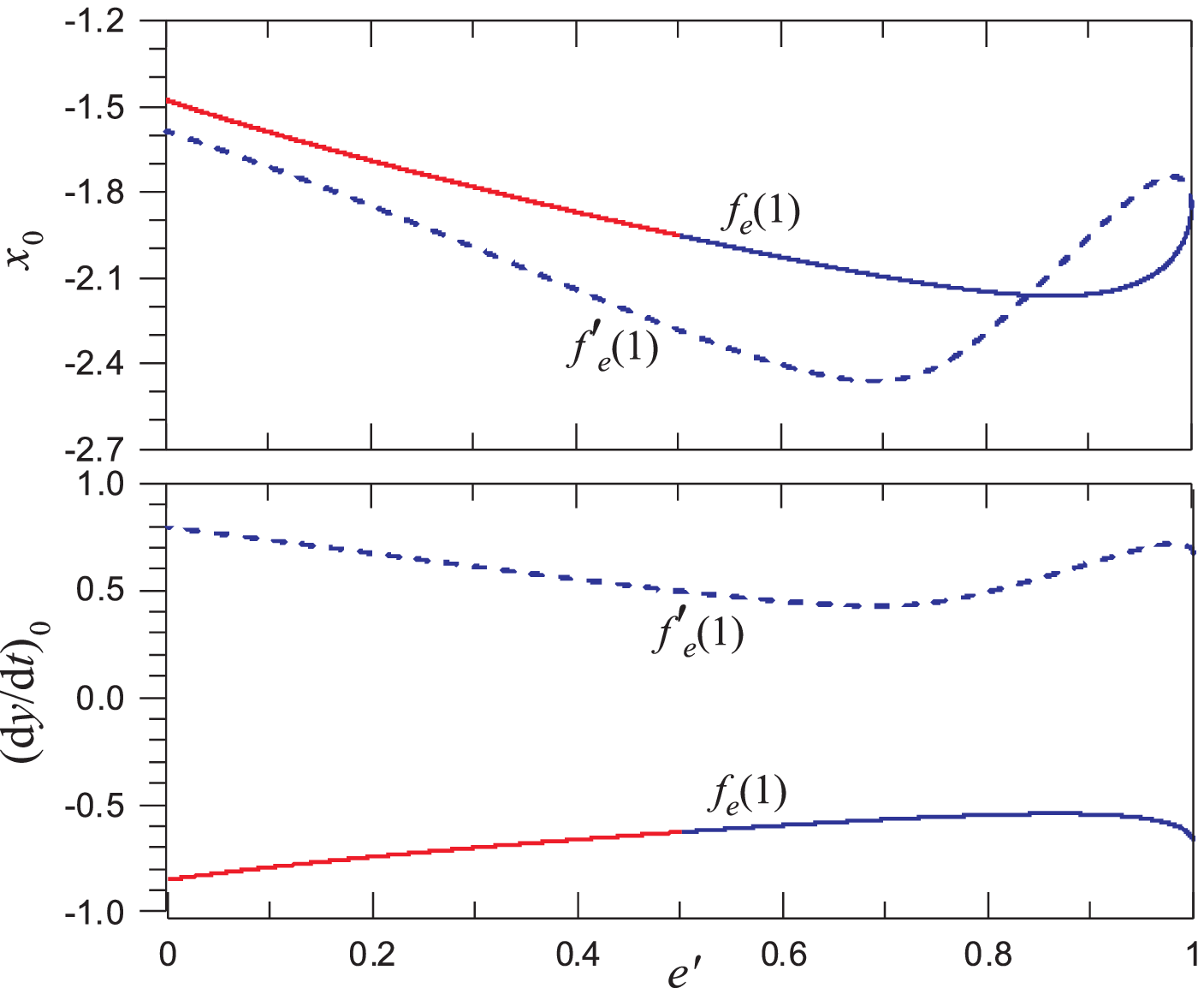} &\quad & \includegraphics[width=0.48\textwidth]{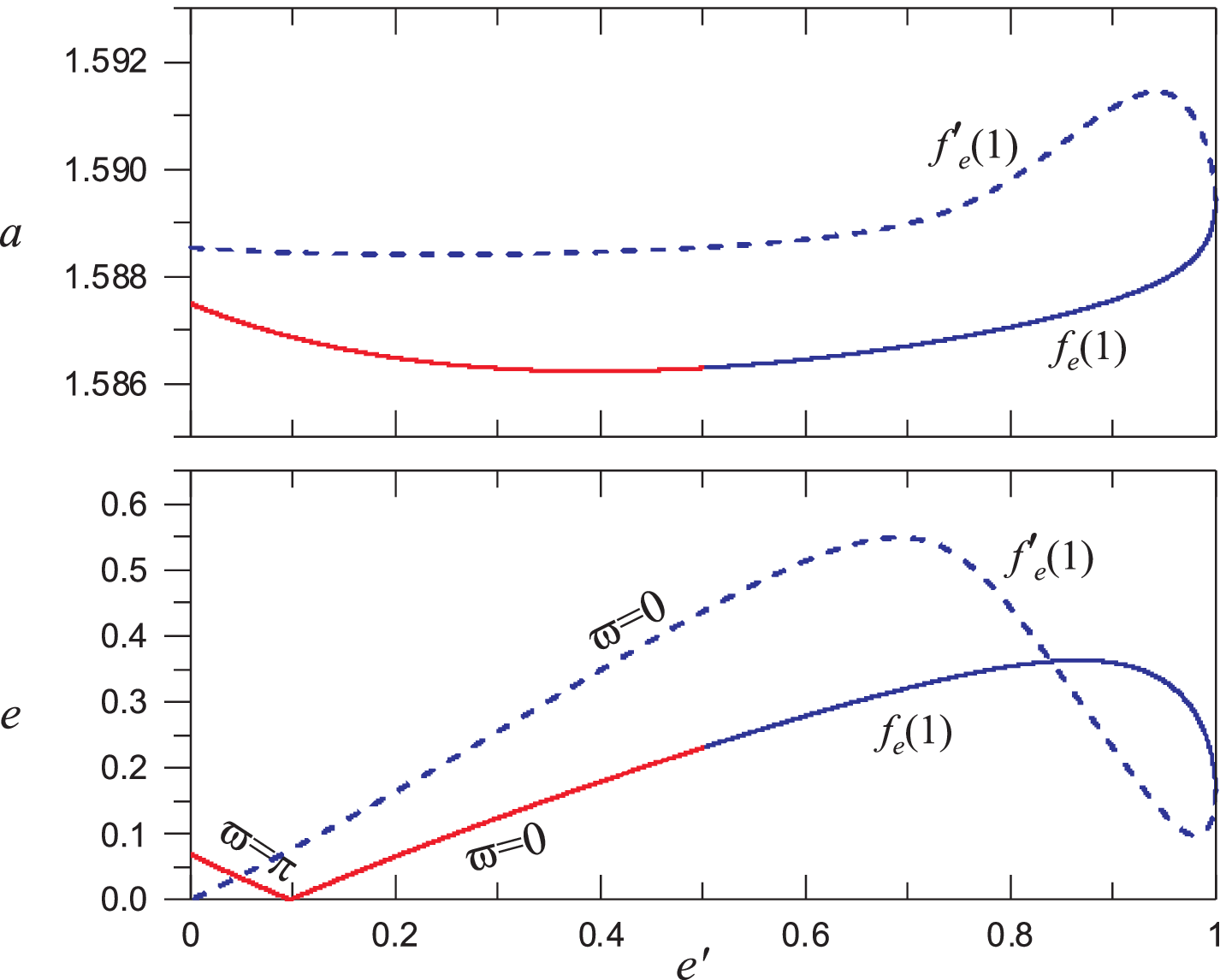} \\
\end{array}$
\caption{Continuation of periodic orbit 1 with respect to the eccentricity $e'$. Both families $f_e(1)$ and $f'_e(1)$ of prograde and retrograde orbits, respectively, are presented. The panels show the initial conditions and the orbital elements of the periodic orbits along the families. The blue (red) line segments indicate that the orbits are linearly stable (single unstable).}
\label{FigConde1}
\end{figure} 

\begin{figure}
$
\begin{array}{ccc}
\includegraphics[width=0.48\textwidth]{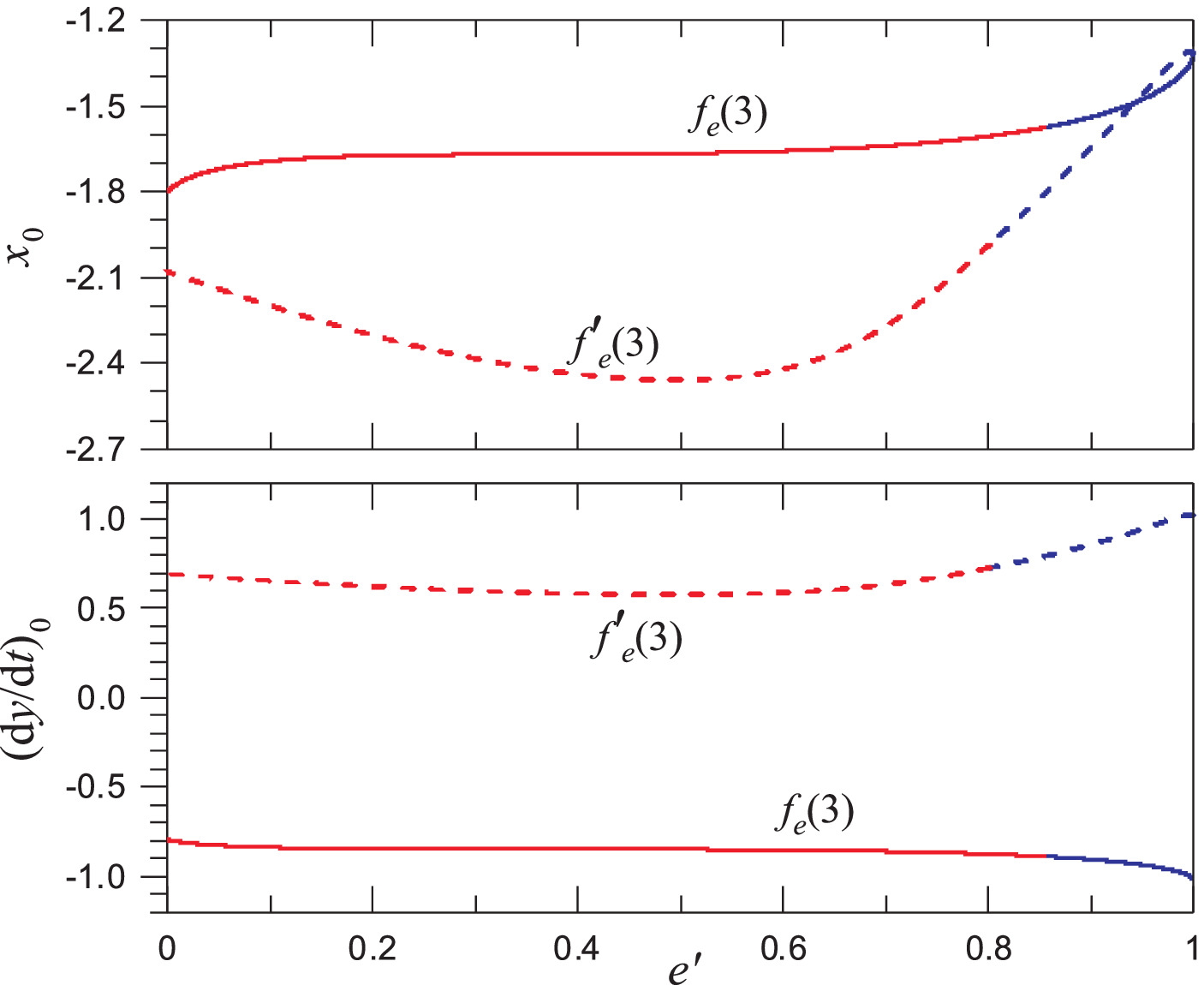} &\quad & \includegraphics[width=0.48\textwidth]{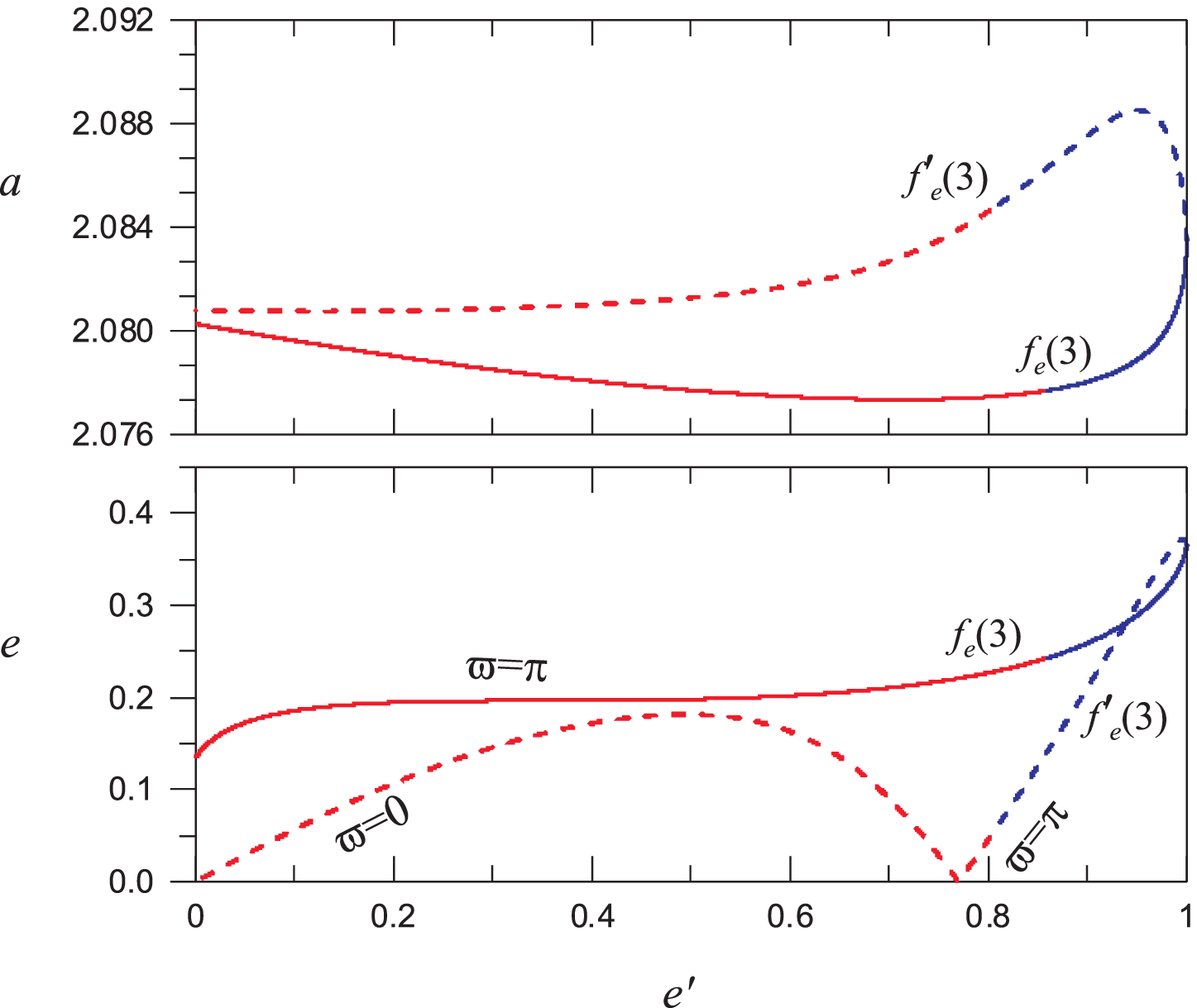} \\
\end{array}$
\caption{Continuation of periodic orbit 3 with respect to the eccentricity $e'$ and for fixed value $\mu=0.001$. Notation is similar to one in Fig. \ref{FigConde1}}
\label{FigConde3}
\end{figure}

\subsection{Continuation with respect to the eccentricity}\label{SectionEcont}
As we have mentioned, due to the symmetry of the rectilinear problem with respect to the $y$-axis, if we change the sign of the $\dot y_0$ we obtain the same periodic orbit. However, for $e'\neq 1$ the sign of $\dot y_0$ determines two different orbits, one prograde and one retrograde. E.g., since in our model the primaries always revolve anti-clockwise, if the small body is located on the right of the massive primary, $x_0>x_{10}$, and $\dot y_0>0$ the orbit is prograde, else is retrograde.  
In the framework of the ERTBP, we computed the families $f_e(1)$ and $f'_e(1)$ for prograde and retrograde orbits, respectively, by fixing the mass value at $\mu=0.001$ and decreasing $e'$. The variation of the initial conditions and orbital elements is shown in Fig. \ref{FigConde1}. Both families are defined in the whole range of eccentricities, $0\leq e'\leq 1$. Family $f'_e(1)$ consists entirely of stable periodic orbits and their eccentricity, $e$, reaches high values around $e'=0.7$ but as $e'\rightarrow 0$ the family ends at a member of the circular family of retrograde orbits of the CRTBP \citep{Henon97}. The family of prograde orbits changes stability at $e'=0.5$. Also at $e'\approx 0.1$ its orbit is circular and the orientation of the family orbits changes from $\varpi=0$ to  $\varpi=\pi$ as $e'$ decreases. The family ends at $e'=0$ to the periodic orbit of period $T=4\pi$ in the $1:2$ resonant family (see e.g. family $II_{(e)}$ in \cite{vkh09}). Similar characteristic curves we obtain for the families generated from orbit 3 and presented in Fig. \ref{FigConde3}. In this case continuation also provides the families $f_{e}(3)$ and $f'_{e}(3)$ in the  whole range of eccentricities, $0\leq e'\leq 1$. Both families start at $e'=1$ with stable orbits, but for lower $e'$ value, the orbits become unstable. The family $f'_{e}(3)$ ends for $e'=0$ at a periodic orbit of period $6\pi$ of the  circular family of retrograde orbits of the CRTBP. The family $f_{e}(3)$ ends at an orbit of the $1:3$ resonant family (with progarde orbits) of the CRTBP.   

In Table \ref{tabs} we present the stability domains obtained from the continuation of the orbits of Table \ref{tab}, both with respect to $\mu$ with fixed value $e'=1$ and with respect to $e'$ with fixed $\mu=0.001$. We can observe that stable periodic orbits exist only for relatively small values of $\mu$. On the other hand continuation with respect to $e'$ provides the families $f'_e(1)$, $f_e(4)$, $f'_e(4)$ and $f'_e(5)$, which are entirely stable. Nevertheless we can notice that for families $f'_e(2)$, $f_e(7)$, $f'_e(7)$, $f_e(8)$ and $f'_e(8)$ the orbits become unstable if we decrease slightly the eccentricity $e'$ from unit. The families may be not continued for all range of parameters. Generally, continuation of families stop at collisions or bifurcation points. In the rectilinear model or in ERTBP there are cases where our computations fail to continue the periodic solutions due to the presence of strong instabilities. In the same table we present also the limit values of  continuation, $\mu_{max}$ and $e'_{min}$.   

\begin{table}
\caption{Stability parameter domains (SPD) after continuation of the stable periodic orbits of table \ref{tab}. For continuation with respect to $e'$, the SPD and $e'_{min}$ values are given for both prograde (families $f_e(n)$) and retrograde orbits (families $f'_e(n)$).} 
{
\begin{tabular}{ccccccc}
\hline
   & stability domain &  & SPD of $f_e(n)$ &  & SPD of $f'_e(n)$ \\ 
$n$ & ($e'=1$) & $\mu_{max}$ & ($\mu=0.001$) & $e'_{min}$ & ($\mu=0.001$) & $e'_{min}$\\ 
\hline
1  & $0\leq \mu<0.0381$ & 0.5  & $0.501\leq e' \leq 1$ & 0  & $0\leq e' \leq 1$ & 0\\ 
   & $0.0744<\mu<0.0873$ &   &  &  &  & \\
2  & $0\leq \mu<0.0010$ & 0.36  & $0.763< e'\leq 1$ & 0.712 & $0.9999<e'\leq 1$ & 0.883 \\ 
3  & $0\leq\mu<0.0081$  & 0.5  & $0.857< e'\leq 1$ & 0 & $0.802< e'\leq 1$ & 0 \\
4  & $0\leq\mu<0.003$   & 0.5 & $0\leq e' \leq 1$ & 0 & $0\leq e' \leq 1$ & 0 \\
5  & $0\leq\mu<0.005$   & 0.5 & $0.744< e'\leq 1$ & 0 & $0\leq e' \leq 1$ & 0\\
6  & $0.00014\leq\mu<0.0045$ & 0.5 & $0.851< e'\leq 1$ & 0  & $0.835< e'\leq 1$ & 0.835 \\
   &   &   & $0.0087< e'\leq 0.0142$ &  &  & \\
7  & $0.0005<\mu<0.0013$ & 0.0025 & $0.9995< e'\leq 1$ & 0.802 & $0.9998< e'\leq 1$ & 0.997 \\
8  & $0.0008<\mu<0.00194$ &  0.0034 & $0.9999< e'\leq 1$ & 0.992 & $0.998<e'\leq 1$ & 0.992\\ 
\hline
\end{tabular}
}
\label{tabs}
\end{table}

\section{An application to the system HD7449}
The extrasolar system HD7449 consists of two massive planets with high eccentricities. Particularly, in normalized units and according to the observations \citep{dumu11},  the inner planet, $P_1$, has $m_1=0.001$, $e_1=0.82\pm0.06$ and $a_1=1$ while the outer planet, $P_2$, has $m_2=0.0019$, $e_2=0.53\pm 0.08$ and $a_2=2.157$. The ratio of revolution period is $T_2/T_1=3.1\pm 0.24$ indicating that the system is possibly captured in the 3:1 mean motion resonance. Also the apsides are rather close to be aligned, $\Delta\varpi=\varpi_2-\varpi_1=32^\circ\pm 14$, instead of anti-aligned.   

\begin{figure}
$
\begin{array}{ccc}
\includegraphics[width=0.48\textwidth]{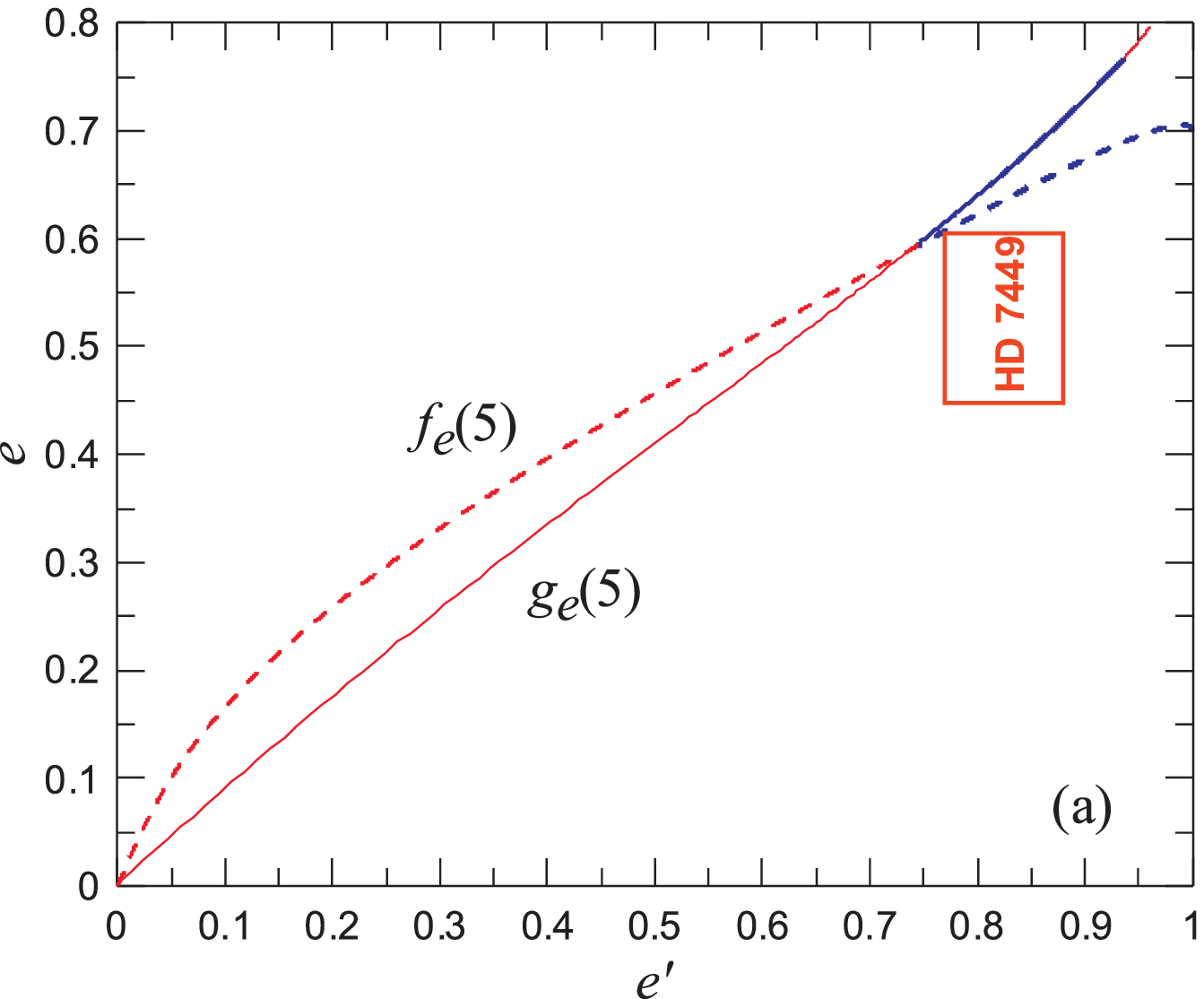}  &\quad & \includegraphics[width=0.48\textwidth]{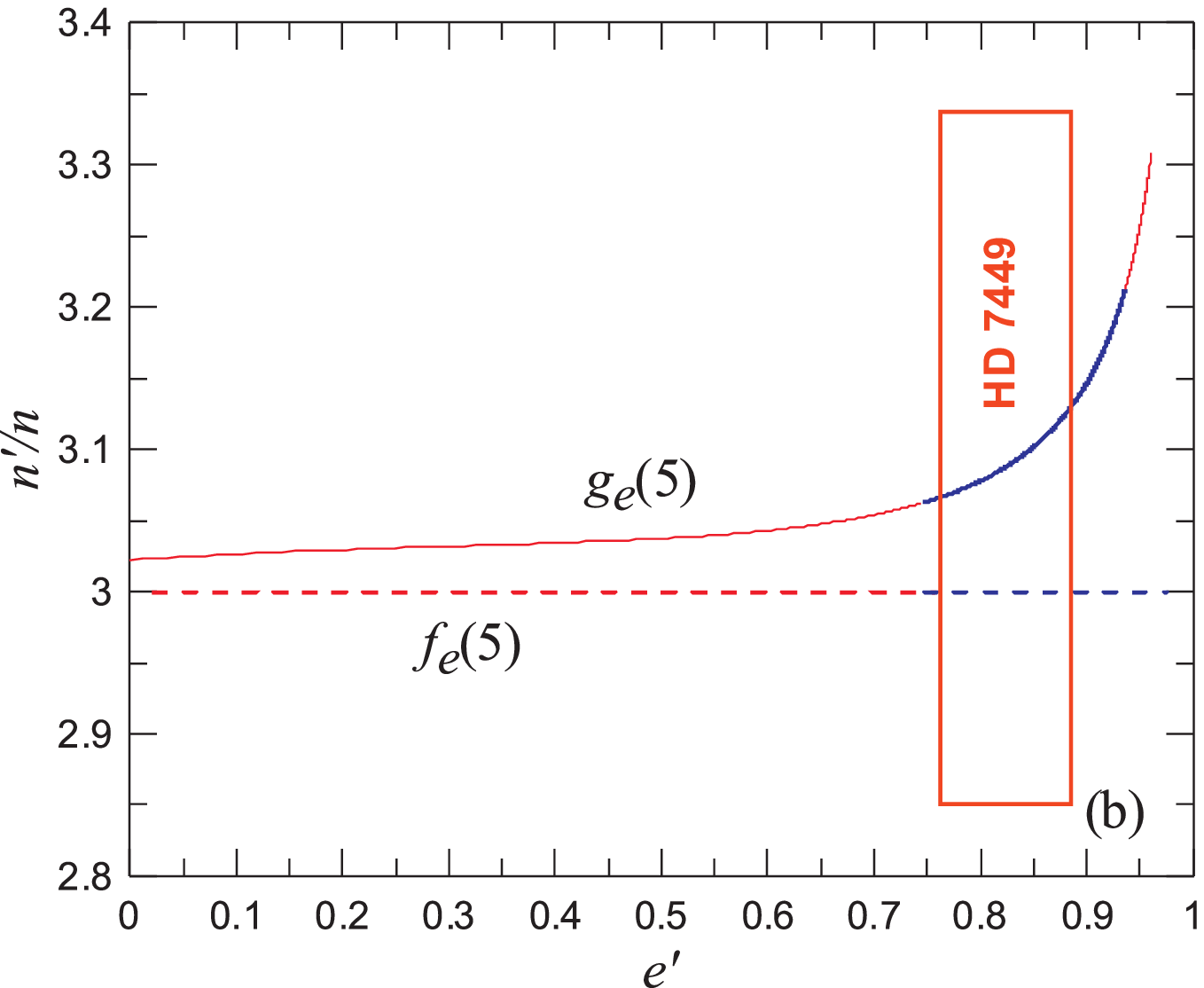} \\
\end{array}$
\caption{{\bf a.} The eccentricity $e$ of the outer planet with respect of the eccentricity $e'$ of the inner planet along the families $f_e(5)$ and $g_e(5)$ {\bf b.} The mean motion ratio of the orbits of the same families. For both planets the rectangle indicates the possible position of the system HD7449 according to the observations. Blue and red segments of the families correspond to linearly stable and unstable orbits, respectively.}
\label{FigFam5}
\end{figure}

From the periodic orbits presented in Table \ref{tab}, the most relative orbit to the HD7449 orbital configuration is orbit 5. In this case we should consider as small primary the inner planet ($\mu=0.001$ and $e'=1$) and continue the periodic orbit by decreasing $e'$. We obtain the family $f_e(5)$ which is presented in Fig. \ref{FigFam5}a. At $e'=e_1=0.82$ the periodic orbit correspond to $e=0.63$, which is slightly larger than the actual value $e_2\approx 0.53$ of the outer planet. Of course all orbits of $f_e(5)$ are of constant period $T=6\pi$ and are exactly 3:1 resonant. Thus, $n/n'\approx 3$ as it is shown in Fig. \ref{FigFam5}b. 

In the above approximation given in the framework of the ERTBP, the outer planet $P_2$ is assumed as a small planet though it is more massive than the inner one. In such cases gravitational interactions among planets are very important for their orbital dynamics and the system should be studied in the framework of the general TBP (GTBP). It can be shown that all periodic orbits of the elliptic restricted problem, where, in our case, the small body is the outer planet $P_2$, can be continued for  nonzero but small values of the mass $m_2$ \citep{HadjiCrist75,ichkm78}. In general, numerical computations show that such a continuation is feasible for large values of $m_2$ \citep{gv17}. 

According to the above mentioned continuation scheme, we can compute periodic solutions for the actual values of the planetary masses of system HD7449. Particularly, we continue the periodic orbits of family $f_e(5)$, which correspond to $m_2=0$, by using the planar GTBP and increasing the mass $m_2$. For $m_2=0.0019$ we obtain the family $g_e(5)$ which is shown in Fig. \ref{FigFam5}. Similarly to family $f_e(5)$, family $g_e(5)$ has a segment of stable orbits at high eccentricities, which is located close to the potential position of the system HD7449. We also observe that the mean motion ratio, $n/n'$, along the family $g_e(5)$ varies and for high eccentricity $e'$ becomes quite larger than the fractional value $3/1$. However, we note that the elliptic periodic orbits of the GTBP are exactly resonant from a dynamical point of view \citep{hadjidem06}. We can observe that the mean value of the mean motion ratio for the planets of HD7449, which is $n'/n\approx 3.1$ is very close to the value indicated by the periodic orbit at $e_1=0.82$.         
  
\begin{figure}
$
\begin{array}{ccc}
\includegraphics[width=0.48\textwidth]{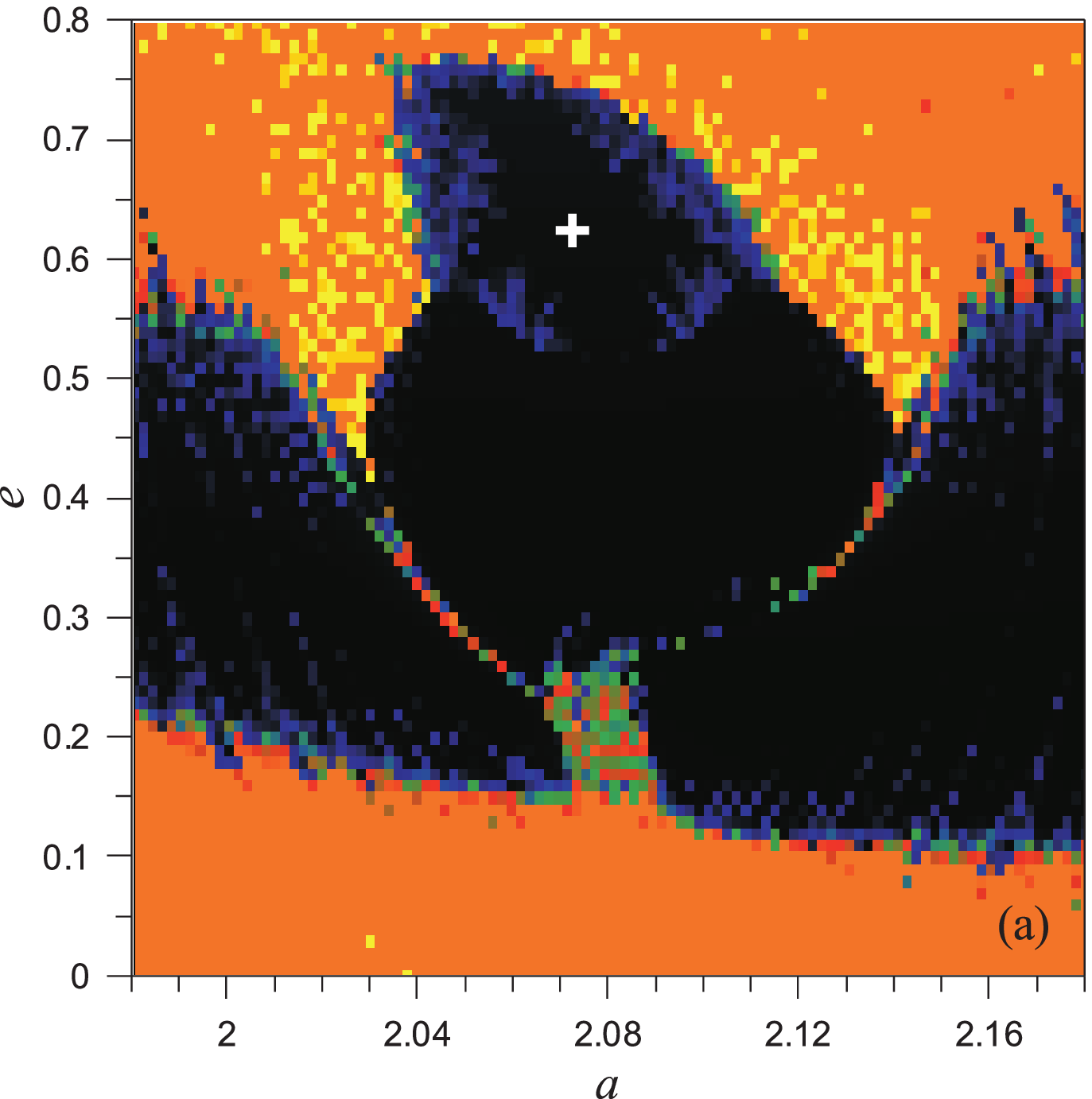} &\quad & \includegraphics[width=0.48\textwidth]{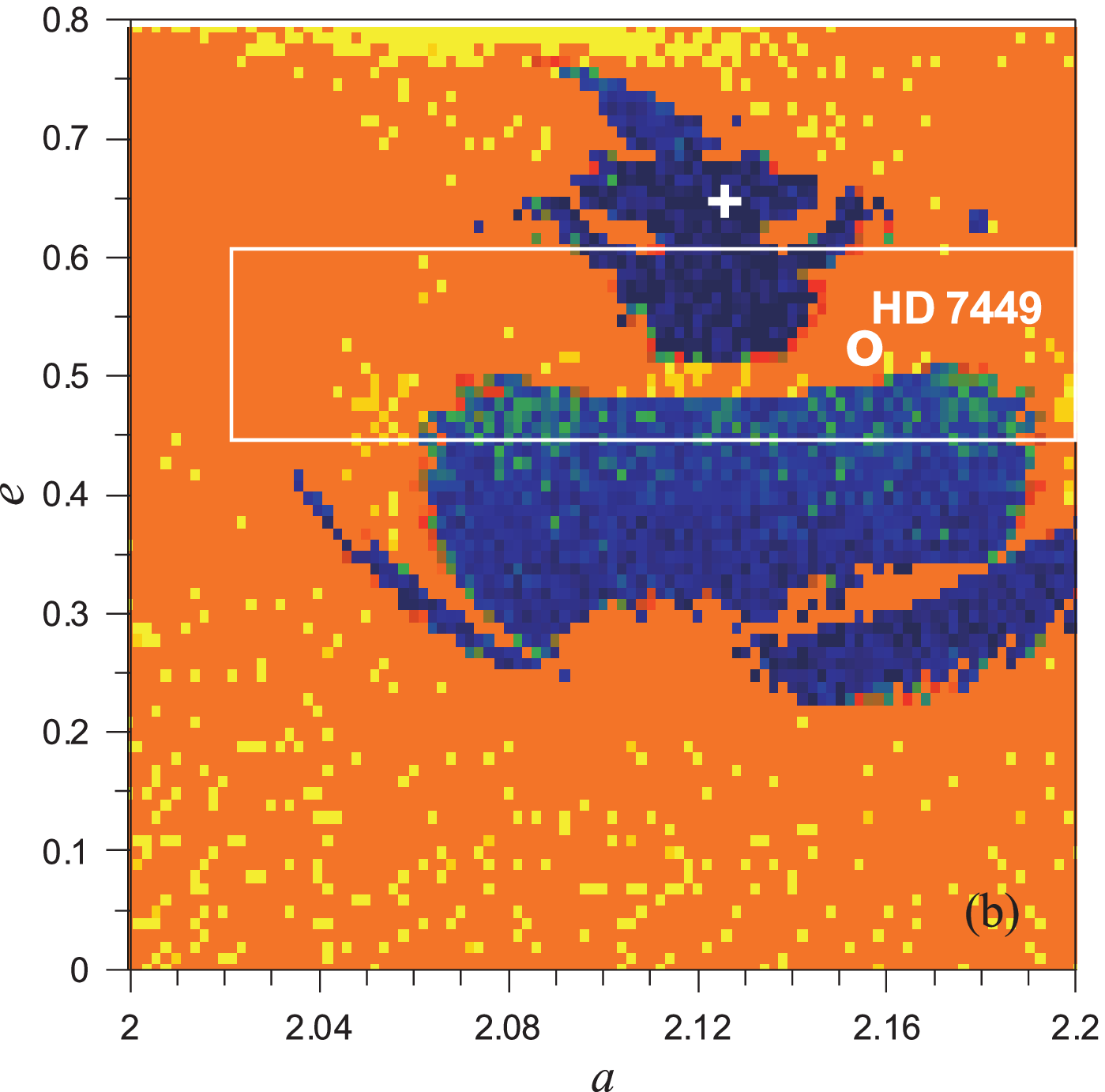} \\
\end{array}$
\caption{Stability maps of size $100\times 80$ with color levels as in Fig. \ref{FigPoMaps} {\bf a.} grid of initial conditions on the plane ($a,e$) for $e'=0.82$ , $\varpi'=M'=0$ in the framework of the ERTBP with $m_1=0.001$, $m_2=0$. The cross indicates the position of the stable periodic orbit of the family $f_e(5)$ {\bf b.} grid of initial conditions as in panel {\bf a} but in the framework of the planar GTBP with $m_1=0.001$, $m_2=0.0019$. The cross indicates the periodic orbit of the family $g_e(5)$. The observations place the system inside a rectangle that intersects our map as it is shown. The circle indicates the position of the system as given in {\em exoplanet.eu} \citep{Schneider11}.}
\label{FigMap5}
\end{figure} 

We consider the normalized semimajor axis of the outer planet, $a$, and its eccentricity, $e$, and in Fig. \ref{FigMap5} we present stability maps on the plane ($a,e$) around the position of the periodic orbit computed for $e'=e_1=0.82$.  In  panel (a) we consider the ERTBP model assuming that $m_2=0$. We can observe a quite familiar picture of a resonant region which is located in the center and is separated from the remaining domain by a thin chaotic separatrix. At high eccentricities, only the domain around the periodic orbit of the family $f_e(5)$ remains to support regular motion. In this region ($e>0.55$) both the angles $\theta_{12}$ and $\varpi$ librate and this is also the case for the resonant angle $\theta=\lambda'-3\lambda+2\varpi$. In the remaining stable regions only $\varpi$ librates around $0^\circ$ and this is a sufficient condition for a well separation between planetary orbits. When we set mass to the outer planet ($m_2=0.0019$) the stability region is strongly affected (see panel (b)). The stable region outside the resonance shrinks but large stability regions survive under the additional perturbation. Now the stability region is separated in a high eccentricity domain, around the periodic orbit of the $g_e(5)$ family, and in a broader region of moderate eccentricities. In the first region the 3:1 resonant angles, $\theta_i=\lambda_1-3\lambda_2+2\varpi_i$, $i=1,2$, and consequently the apsidal difference $\Delta\varpi$, librate around $0^\circ$. In the region of moderate eccentricities only $\Delta\varpi$ librates. The position of the HD7449, given from the mean value of the analysis of observations is located in the chaotic regime. However, the area of possible location of the HD7449, according to the error-bars in the observations, includes parts of both regions of stability. Hence, if $\theta_i$ librate, the system should be located in the high eccentricity regime of stability around the periodic orbit. If only $\Delta\varpi$ librates then the orbit of the outer planet should be of moderate eccentricity ($0.25<e_2<0.5$).  In \citet{av16} the orbital stability of HD7449 is studied with respect to the families of periodic orbits of the GTBP and asymmetric configurations are also included. The presented analysis emphasizes more clearly the broadness and the distinction of the stability domains around the particular symmetric periodic orbit of the family $g_e(5)$, which coincides with the family $S_3$ of the above mentioned paper.       

\section{Conclusions}
In this paper we study the orbital dynamics of the rectilinear elliptic restricted TBP. In a way similar to using the circular restricted TBP for probing the dynamics at small or moderate eccentricities of the primaries, we show that the rectilinear elliptic restricted TBP can be used for understanding the stability of motion when the primaries move on highly eccentric orbits. 

The backbone of the dynamics of the rectilinear model is its set of isolated periodic orbits. A first study of symmetric periodic orbits has been given by Broucke (in paper-I). We studied the Broucke's orbits, with respect to their linear stability, and all of them found to be unstable. Such instability seems to form very unstable regions in phase space which cause the fast escape of the small body. Nevertheless, region of stable orbits are revealed, by constructing escape-time or FLI maps, which exist when the small body revolves in an orbit far from the primaries and the system is described by a hierarchical orbital configuration. However, stability regions also exist when the gravitational interactions of both primaries are quite strong. Such regions correspond to stable resonant motion which takes place around linearly stable periodic orbits.

We performed an extensive numerical search of periodic orbits with period $T=k\,T'$, where $T'$ is the period of the primaries, $k=1,2,3,4$, and for mass parameter $0.001\leq\mu\leq 0.5$. The majority of periodic orbits found are linearly unstable. For $\mu=0.001$, eight stable periodic orbits found, which are given in Table \ref{tab}. Six of them correspond to circumstellar orbits with revolution approximate period $T_r>T'$, two are circumbinary orbits and two are satellite orbits  around the massive primary with $T_r<T'$. We continued these orbits with respect to the mass parameter, $\mu$, for $e'=1$. Such a continuation provided in some cases periodic solutions up to $\mu=0.5$ but their linear stability changes from stability to instability for $\mu>\mu_{max}$.  We performed also continuation with respect to the eccentricity of the primaries, $e'$, and we obtained periodic orbits in the elliptic restricted problem with $e'<1$. From each periodic orbit for $e'=1$ two families bifurcate for $e'<1$, one consists of prograde orbits and one of retrograde orbits. The stability domains determined by our linear analysis are presented in Table \ref{tabs}. 

Stable periodic orbits are surrounded in phase space with invariant tori and guarantee the long term stability of orbits. Our study showed the existence of significant stability regions even when the primaries revolve in very high eccentric motion ($e'\rightarrow 1$). A part of such stable regions seems to persist when we set $e'<1$ and nonzero mass to the small body i.e. when we continue the periodic orbits of the rectilinear problem to the ERTBP and then to the GTBP. Therefore, stable configurations for massive planets on highly eccentric orbits, can be found in this way. We showed that such a stable configuration is related with the HD7449 extrasolar system, which is an example of a real system consisting of two very eccentric planetary companions.         

\bibliographystyle{plainnat}
\bibliography{gvnbib}

\begin{thebibliography}{21}
\providecommand{\natexlab}[1]{#1}
\providecommand{\url}[1]{\texttt{#1}}
\expandafter\ifx\csname urlstyle\endcsname\relax
  \providecommand{\doi}[1]{doi: #1}\else
  \providecommand{\doi}{doi: \begingroup \urlstyle{rm}\Url}\fi

\bibitem[{Antoniadou} and {Voyatzis}(2016)]{av16}
K.~I. {Antoniadou} and G.~{Voyatzis}.
\newblock {Orbital stability of coplanar two-planet exosystems with high
  eccentricities}.
\newblock \emph{Monthly Notices of the Royal Astronomical Society},
  461:\penalty0 3822--3834, 2016.

\bibitem[{Barnes} and {Greenberg}(2006)]{barnes06}
R.~{Barnes} and R.~{Greenberg}.
\newblock {Stability Limits in Extrasolar Planetary Systems}.
\newblock \emph{The Astrophysical Journal}, 647:\penalty0 L163--L166, 2006.
\newblock \doi{10.1086/507521}.

\bibitem[Broucke(1969)]{broucke69}
R.~Broucke.
\newblock Stability of periodic orbits in the elliptic, restricted three-body
  problem.
\newblock \emph{AIAA Journal}, 7:\penalty0 1003--1009, 1969.

\bibitem[{Contopoulos} and {Harsoula}(2012)]{Contopoulos12}
G.~{Contopoulos} and M.~{Harsoula}.
\newblock {Chaotic spiral galaxies}.
\newblock \emph{Celestial Mechanics and Dynamical Astronomy}, 113:\penalty0
  81--94, 2012.
\newblock \doi{10.1007/s10569-011-9378-7}.

\bibitem[Danby(1987)]{danby87}
J.~M.~A. Danby.
\newblock The solution of kepler's equation.
\newblock \emph{Celestial Mechanics}, 40:\penalty0 303--312, 1987.

\bibitem[{de Assis} and {Terra}(2014)]{Assis14}
S.~C. {de Assis} and M.~O. {Terra}.
\newblock {Escape dynamics and fractal basin boundaries in the planar
  Earth-Moon system}.
\newblock \emph{Celestial Mechanics and Dynamical Astronomy}, 120:\penalty0
  105--130, 2014.
\newblock \doi{10.1007/s10569-014-9567-2}.

\bibitem[{Dumusque} et~al.(2011){Dumusque}, {Lovis}, {S{\'e}gransan}, {Mayor},
  {Udry}, {Benz}, {Bouchy}, {Lo Curto}, {Mordasini}, {Pepe}, {Queloz},
  {Santos}, and {Naef}]{dumu11}
X.~{Dumusque}, C.~{Lovis}, D.~{S{\'e}gransan}, M.~{Mayor}, S.~{Udry},
  W.~{Benz}, F.~{Bouchy}, G.~{Lo Curto}, C.~{Mordasini}, F.~{Pepe},
  D.~{Queloz}, N.~C. {Santos}, and D.~{Naef}.
\newblock {The HARPS search for southern extra-solar planets. XXX. Planetary
  systems around stars with solar-like magnetic cycles and short-term activity
  variation}.
\newblock \emph{Astronomy \& Astrophysics}, 535:\penalty0 A55, 2011.

\bibitem[Froeschl{\'e} et~al.(1997)Froeschl{\'e}, Lega, and Gonczi]{froe97}
C.~Froeschl{\'e}, E.~Lega, and R.~Gonczi.
\newblock Fast lyapunov indicators. application to asteroidal motion.
\newblock \emph{Celestial Mechanics and Dynamical Astronomy}, 67:\penalty0
  41--62, 1997.

\bibitem[Hadjidemetriou(2006)]{hadjidem06}
J.~D. Hadjidemetriou.
\newblock Symmetric and asymmetric librations in extrasolar planetary systems:
  a global view.
\newblock \emph{Celestial Mechanics and Dynamical Astronomy}, 95:\penalty0
  225--244, 2006.

\bibitem[{Hadjidemetriou} and {Christides}(1975)]{HadjiCrist75}
J.~D. {Hadjidemetriou} and T.~{Christides}.
\newblock {Families of periodic orbits in the planar three-body problem}.
\newblock \emph{Celestial Mechanics}, 12:\penalty0 175--187, 1975.

\bibitem[{Henon}(1997)]{Henon97}
M.~{Henon}.
\newblock \emph{{Generating Families in the Restricted Three-Body Problem}}.
\newblock Springer-Verlag, 1997.

\bibitem[Ichtiaroglou et~al.(1978)Ichtiaroglou, Katopodis, and
  Michalodimitrakis]{ichkm78}
S.~Ichtiaroglou, K.~Katopodis, and M.~Michalodimitrakis.
\newblock On the continuation of periodic orbits in the three-body problem.
\newblock \emph{Astronomy and Astrophysics}, 70:\penalty0 531, 1978.

\bibitem[{Kov{\'a}cs} and {{\'E}rdi}(2009)]{Kovacks09}
T.~{Kov{\'a}cs} and B.~{{\'E}rdi}.
\newblock {Transient chaos in the Sitnikov problem}.
\newblock \emph{Celestial Mechanics and Dynamical Astronomy}, 105:\penalty0
  289--304, 2009.
\newblock \doi{10.1007/s10569-009-9227-0}.

\bibitem[{P\'aez} and {Efthymiopoulos}(2015)]{Paez2015}
R.~I. {P\'aez} and C.~{Efthymiopoulos}.
\newblock {Trojan resonant dynamics, stability and chaotic diffusion, for
  parameters relevant to exoplanetary systems}.
\newblock \emph{Celestial Mechanics and Dynamical Astronomy}, 121:\penalty0
  139--170, 2015.
\newblock \doi{10.1007/s10569-014-9591-2}.

\bibitem[{Pichardo} et~al.(2008){Pichardo}, {Sparke}, and
  {Aguilar}]{pichardo08}
B.~{Pichardo}, L.~S. {Sparke}, and L.~A. {Aguilar}.
\newblock {Geometrical and physical properties of circumbinary discs in
  eccentric stellar binaries}.
\newblock \emph{Monthly Notices of the Royal Astronomical Society},
  391:\penalty0 815--824, 2008.

\bibitem[{Pilat-Lohinger} and {Dvorak}(2002)]{dvorak02}
E.~{Pilat-Lohinger} and R.~{Dvorak}.
\newblock {Stability of S-type Orbits in Binaries}.
\newblock \emph{Celestial Mechanics and Dynamical Astronomy}, 82:\penalty0
  143--153, 2002.

\bibitem[{Schneider} et~al.(2011){Schneider}, {Dedieu}, {Le Sidaner},
  {Savalle}, and {Zolotukhin}]{Schneider11}
J.~{Schneider}, C.~{Dedieu}, P.~{Le Sidaner}, R.~{Savalle}, and
  I.~{Zolotukhin}.
\newblock {Defining and cataloging exoplanets: the exoplanet.eu database}.
\newblock \emph{Astronomy \& Astrophysics}, 532:\penalty0 A79, 2011.

\bibitem[{Schubart}(1956)]{schubart56}
J.~{Schubart}.
\newblock {Numerische aufsuchung periodischer Losungen im dreikorper-problem}.
\newblock \emph{Astron. Nachr.}, 283:\penalty0 17--22, 1956.

\bibitem[{Schubart}(2017)]{schubart17}
J.~{Schubart}.
\newblock Libration of arguments of circumbinary-planet orbits at resonance.
\newblock \emph{Celestial Mechanics and Dynamical Astronomy}, 2017.
\newblock \doi{10.1007/s10569-017-9752-1}.

\bibitem[{Voyatzis}(2017)]{gv17}
G.~{Voyatzis}.
\newblock Periodic orbits of planets in binary systems.
\newblock In T.~A. {Maindl}, H.~{Varvoglis}, and R~{Dvorak}, editors,
  \emph{Proccedings of the first Greek-Austrian workshop on extrasolar
  planetary systems}, 2017.
\newblock ISBN 10 1544255632.

\bibitem[Voyatzis et~al.(2009)Voyatzis, Kotoulas, and Hadjidemetriou]{vkh09}
G.~Voyatzis, T.~Kotoulas, and J.~D. Hadjidemetriou.
\newblock On the 2/1 resonant planetary dynamics - periodic orbits and
  dynamical stability.
\newblock \emph{Monthly Notices of the Royal Astronomical Society},
  395:\penalty0 2147--2156, 2009.

\end{thebibliography}
\end{document}